\documentclass[aps,prd,preprint,preprintnumbers,showpacs]{revtex4-1}
\usepackage{graphicx}
\usepackage{amsmath,amsfonts,amssymb}
\usepackage[section] {placeins}
\usepackage{multirow}

\begin{document}

\title{On the absorption and production cross sections of  $K$ and $K^*$}
\author{A.~Mart\'inez~Torres\footnote{amartine@if.usp.br}}

 \affiliation{
Instituto de F\'isica, Universidade de S\~ao Paulo, C.P. 66318, 05389-970 S\~ao 
Paulo, SP, Brazil.
}
\author{K.~P.~Khemchandani\footnote{kanchan@if.usp.br}}
 \affiliation{
Universidade Federal de S\~ao Paulo, C.P. 01302-907, S\~ao Paulo, Brazil
}

\author{ L. M. Abreu\footnote{luciano.abreu@ufba.br}}   
\affiliation{Instituto de F\'isica, Universidade Federal da Bahia, 
Campus Universit\'ario de Ondina, 40170-115, Bahia, Brazil}

\author{ F.~S.~Navarra\footnote{navarra@if.usp.br}}
 \affiliation{
Instituto de F\'isica, Universidade de S\~ao Paulo, C.P. 66318, 05389-970 S\~ao 
Paulo, S\~ao Paulo, Brazil.
}
\author{ M.~Nielsen\footnote{mnielsen@if.usp.br} }
 \affiliation{
Instituto de F\'isica, Universidade de S\~ao Paulo, C.P. 66318, 05389-970 S\~ao 
Paulo, S\~ao Paulo, Brazil.
}

\preprint{}

\date{\today}

\begin{abstract}
We have computed the isospin and spin averaged cross sections of the  
processes $\pi K^*\to \rho K$ and $\rho K^*\to \pi K$, which are crucial in 
the determination of the abundances  of $K^*$ and $K$ 
in heavy ion collisions. Improving previous calculations, we have considered 
several mechanisms which were missing, such as the exchange  
of  axial and vector resonances ($K_1(1270)$, $K^*_2(1430)$, $h_1(1170)$, 
etc...)  and also other processes such as  $\pi K^*\to \omega K, \phi K$ and 
$\omega K^*,\,\phi K^*\to \pi K$. We find that some of these mechanisms give important 
contributions to the cross section.  Our results also suggest that, in a hadron gas, 
$K^*$ production might be more important than its absorption. 
\end{abstract}


\maketitle
 
\section{Introduction}

The study of nucleus-nucleus collisions at high 
energies~\cite{star1,star2,alice1,alice2}, such as Au+Au  at center of mass 
energies of 200 GeV or Pb-Pb at center of mass energies of 2.76 TeV, hints towards 
the existence of a phase transition from nuclear matter to a locally thermalized 
state of deconfined quarks and gluons, the quark-gluon plasma (QGP)~\cite{blum}. 
After a hot initial stage, the QGP cools and hadronizes forming a hadron gas, where the 
produced mesons and baryons interact inelastically and the relative abundances are 
changed. After further cooling, the system reaches chemical equilibrium, where only 
elastic collisions take place. This is also called ``chemical freeze-out" and at this point the 
abundances are frozen.  Finally, at the ``kinetic freeze-out", the density becomes small, 
the interactions no longer occur and the particles stream freely to the detectors 
\cite{song,rapp}. After hadronization  and before the kinetic freeze-out the hadrons can interact and  
different production and absorption reactions (including the formation and 
decay of resonances) will change the  hadron abundances. These changes will be different 
for different hadron species and they depend on the details of hadron dynamics, especially 
on possible resonance formation.

Particularly interesting is the case of the $K^*(892)$ meson. The lifetime of this 
meson is around 4 fm/c, which is smaller than that of the QGP formed in heavy-ion 
collisions ($\sim 10$ fm/c~\cite{rapp}). This means that, from hadronization up to 
the kinetic freeze-out,  a $K^*$ meson present in the QGP has 
enough time to decay into $K$ and $\pi$.  It can also be absorbed, as well as produced, 
by other mesons present in the medium. All these reactions can change the abundance of 
the $K^*$ at the kinetic freeze-out. 

In Refs.~\cite{star1,star2,alice1,alice2}, $K^*$ production was investigated  
considering data from Au$+$Au at center of mass energies of 200 GeV, from Cu$+$Cu  
at 62.4 and 200 GeV and from Pb$+$Pb collisions at $2.76$ GeV. Considering  the  
$K^*$ and $K$ transverse momentum spectra and the measured $K^*/K$ yield ratios 
for all centralities in Au$+$Au or Cu$+$Cu compared to the same ratio from $p+p$ 
collisions, a significant reduction in the $K^*/K$ ratio  was found. The measured 
values were $0.23\pm 0.05$ in Au$+$Au collisions at 200 GeV at RHIC~\cite{star1} 
and of $0.19\pm0.05$ in Pb$+$Pb collisions at 2.76 TeV at LHC~\cite{alice2}, while 
the statistical model predicts $0.33\pm 0.01$ in case of Au$+$Au collisions at 130 
GeV at RHIC~\cite{braun2}. In all these collisions mesons are produced at hadronization, 
i.e. when quarks and gluons are converted into hadrons, or later, during hadronic 
scatterings in the hadron gas. Any special feature observed in the measured yields reflects 
what happens in these two stages.

In Ref.~\cite{cho}, the changes in the $K^*$ and $K$ abundances caused by hadronic 
scatterings in the hadron gas phase were studied. The authors calculated the 
cross sections for  absorption (and production) of $K$ and $K^*$  by $K$, $K^*$, 
$\pi$ and $\rho$. The following processes were considered to account for $K$ 
absorption: $ K\bar K \to \pi \pi$, $K\bar K\to\rho\rho$ and $K\pi\to K^*$.  
Similarly the absorption of $K^*$ mesons was attributed to the processes: 
$K^*\pi\to K\rho$, $K^*\rho\to K\pi$, 
$K^*\bar K\to\rho\pi$, 
$K^*\bar K^*\to \pi\pi$ and 
$K^*\bar K^*\to\rho\rho$. 
The corresponding production mechanism for $K^*$ and $K$ are simply 
the inverse reactions of those mentioned above, whose cross sections can be 
obtained by using the detailed balance principle. These production and absorption 
cross sections are the most important  input entering in the rate equations through  
which the time evolution of the abundance of both $K^*$ and $K$,  can be obtained. 
As shown in Ref.~\cite{cho}, due to the interactions of $K$ 
and $K^*$ with the hadrons present in the medium, the yield associated with the ratio 
$K^*/K$ decreases by 36\% during the expansion 
of the hadronic matter. The main mechanisms contributing to this  reduction were  
found   to be the processes $K^*\pi\to K\rho$, 
$K^*\rho\to K\pi$, $K^*\to K\pi$ (the corresponding inverse reactions were, of course, 
also included in the calculation). Considering 
these processes, an abundance ratio comparable to the RHIC and LHC measurements was  
found and it was concluded that the measured  ratio  $K^*/K$  can be explained by 
the interaction of $K^*$ and $K$ with light mesons in the hadronic medium. 

In the determination of the cross sections for the reactions $K^*\pi\to K\rho$ 
and $K^*\rho\to K\pi$ performed in  Ref.~\cite{cho}  some mechanisms were ignored and 
they could be relevant for the calculation of the $K^*/K$ ratio  such as, 
for instance, the exchange of axial resonances. To consider resonance exchange, 
though, one needs a reliable information on the mass and width of the resonance 
as well as the couplings at different resonance-meson-meson vertices. Such 
information is available in the literature. For example, it was shown in 
Refs.~\cite{rocaS,gengR} that the $K^*\pi$ interaction and coupled channels 
($\phi K$, $\omega K$, $\rho K$ and $K^*\eta$) generate the 
axial vector meson $K_1(1270)$ state with a two pole structure. The presence of 
this resonances has been found to be important~\cite{gengR} in describing the 
invariant mass distribution of the process $K^-p\to K^-\pi^+\pi^- p$ at 63 GeV  measured  
by the WA3 collaboration at CERN~\cite{daum}. Similarly, the exchange of $K_1(1270)$ 
could also play an important role when determining the cross section of the reaction 
$\pi K^*\to \rho K$.  Reference~\cite{rocaS} also discusses the interaction of $\bar K^* K$ 
and coupled 
channels in different isospin $I$ and $G$-parity combinations, which give rise to 
the following axial resonances listed by the Particle Data Group (PDG)~\cite{PDG}:  
$h_1(1170)$, $h_1(1380)$ for $I=0$, $G=-1$; $f_1(1285)$ for $I=0$, $G=+1$; 
$a_1(1260)$ for $I=1$, $G=-1$; $b_1(1235)$ for $I=1$, $G=+1$. The nature of these 
resonances has been tested in Refs.~\cite{rocaH, francesca,francesca2} where their   
decay widths in several channels were calculated and a good 
description of the experimental data was found. The inclusion of these resonances  
can contribute to the cross section of $\rho K^*\to \pi K$.

The main purpose of the present work is to include 
the exchange of all these resonances in the study of the processes 
$K^*\pi\to K\rho$ and $K^*\rho\to K\pi$. Besides resonance exchange, some other 
mechanisms are missing in the determination of the cross sections of 
$K^* \pi \to K \rho$ and  $K^*\rho\to K\pi$ performed in  Ref.~\cite{cho}. 
For example, the exchange of a vector meson in the $t$-channel and a pseudoscalar 
in the $s$-channel were taken into account to study  the reaction 
$K^* \pi \to K \rho$, but other mechanisms like $u$-channel exchange 
or $s$-channel exchange of vectors were not. Some of such missing 
diagrams involve anomalous vertices~\cite{wess,witten} (i.e., the natural parity 
is not conserved in the vertex, which is described by a Lagrangian containing 
the Levi-Civita pseudotensor). In Refs.~\cite{oh,xprod1} it was shown that 
interaction terms with anomalous parity couplings have a strong impact on the 
corresponding cross sections, and the relevance of such anomalous terms in the 
determination of the abundance of $X(3872)$ in heavy ion collisions was computed 
in Ref.~\cite{xprod2}. Such processes, involving the anomalous vertices, were missed in the earlier work of Ref.~\cite{cho2}.
Similar is the case of the reaction $K^* \rho \to \pi K$: in 
Ref.~\cite{cho} Feynman diagrams related to the exchange of a pseudoscalar meson  
in the $t$-channel and a vector meson in the $s$-channel were considered. However, 
other contributions, as $u$-channel exchange diagrams and exchange of other mesons 
in the $t$- and $s$-channels were not taken into account. In this work we are going 
to evaluate the contribution from all such mechanisms and calculate the cross 
sections of the reactions  $K^*\pi\to K\rho, \, K\omega, K\phi$ and 
$K^*\rho,\, K^*\omega,\, K^*\phi\to K\pi$ for the absorption of the $K^*$ meson and 
the corresponding cross section for its production.

\section{Formalism}

In the model of Ref.~\cite{cho}, the effect of  absorption and production 
of $K^*$ and $K$ mesons in a hadron gas appears in the thermal average 
cross sections of such processes. These cross sections affect the time evolution 
of the abundance of $K^*$ and $K$. As concluded in 
Ref.~\cite{cho}, the most important absorption and production processes of $K^*$ 
and $K$  correspond to $\pi K^*\to \rho K$, $\rho K^*\to \pi K$, $K^*\to\pi K$ and  
the inverse reactions.

In the present work, we calculate these cross sections including the  
following reactions $\pi K^*\to \rho K,\, \omega K,\, \phi K$ and 
$\rho K^*,\, \omega K^*,\, \phi K^*\to\, \pi K$. The cross sections associated 
with the corresponding inverse reactions can be obtained using the principle of 
detailed balance. Note that in Ref.~\cite{cho}, the cross sections related to 
processes involving $\omega$ and $\phi$ in the initial or final states were not 
evaluated in spite of their mass similarity with $\rho$ as well as similar 
dynamics involved in the corresponding reactions.

We will calculate the cross section of the process $a+b\to c+d$. For a specific reaction mechanism  $r$, we can 
write $\sigma_r$ in the center of mass frame as~\cite{cho,xprod1,cho2}
\begin{align}
\sigma_r(s)=\frac{1}{16\pi\lambda(s,m^2_{a,r},m^2_{b,r})}
\int^{t_{\textrm{max,r}}}_{t_\textrm{min,r}}dt\overline{\sum\limits_{S,I}}
\left |\mathcal{M}_r(s,t)\right|^2,\label{cross}
\end{align}
where $s$ and $t$ are the Mandelstam variables for the reaction $r$, $m_{a,r}$ 
and $m_{b,r}$ represent the masses of the two  
particles in the initial state of the reaction $r$, $\lambda(a,b,c)$ is the K\"all\'en 
function and $\mathcal{M}_r$ is the reduced matrix element for the 
process $r$.  

The symbol $\overline{\sum\limits_{S, I}}$ in Eq.~(\ref{cross}) represents the sum over the 
spins ($S$) and isospins ($I$) projections of the particles in the initial and 
final states, weighted by the isospin and spin degeneracy factors of the two 
particles forming the initial state for the reaction $r$, i.e.,
\begin{align}
\overline{\sum\limits_{S,I}}\left|\mathcal{M}_r\right|^2\to           
\frac{1}{(2I_{a,r}+1)(2I_{b,r}+1)}\frac{1}{(2s_{a,r}+1)(2s_{b,r}+1)}
\sum\limits_{S,I}\left|\mathcal{M}_r\right|^2,
\end{align}
where,
\begin{align}
\sum\limits_{S,I}\left|\mathcal{M}_r\right|^2=\sum\limits_{i,j}
\left[\sum\limits_{S}\left|\mathcal{M}^{ij}\right|^2\right].\label{Mqq}
\end{align}
In Eq.~(\ref{Mqq}), $i$ and $j$ represent the initial ($a+b$) and final $(c+d)$ 
channels in the reaction $r$ for a particular charge 
$Q_r=Q_{a}+Q_{b}=Q_c+Q_d=-1,0,+1,+2$.  

\begin{figure}[h!]
\includegraphics[width=0.8\textwidth]{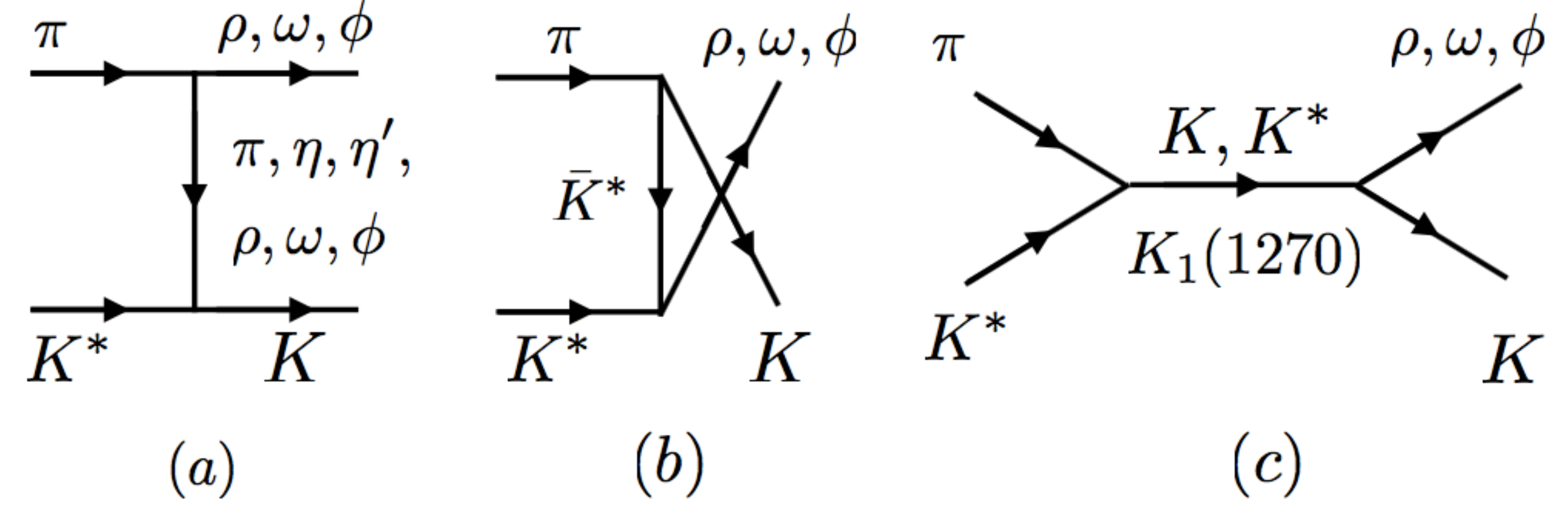}                       
\caption{Diagrams contributing to the processes                                  
$\pi K^*\to \rho K,\, \omega K,\,\phi K$ in the $t$-channel (a), $u$-channel (b) 
and $s$-channel (c).}\label{piKstar}
\end{figure}

\begin{figure}[h!]
\includegraphics[width=0.8\textwidth]{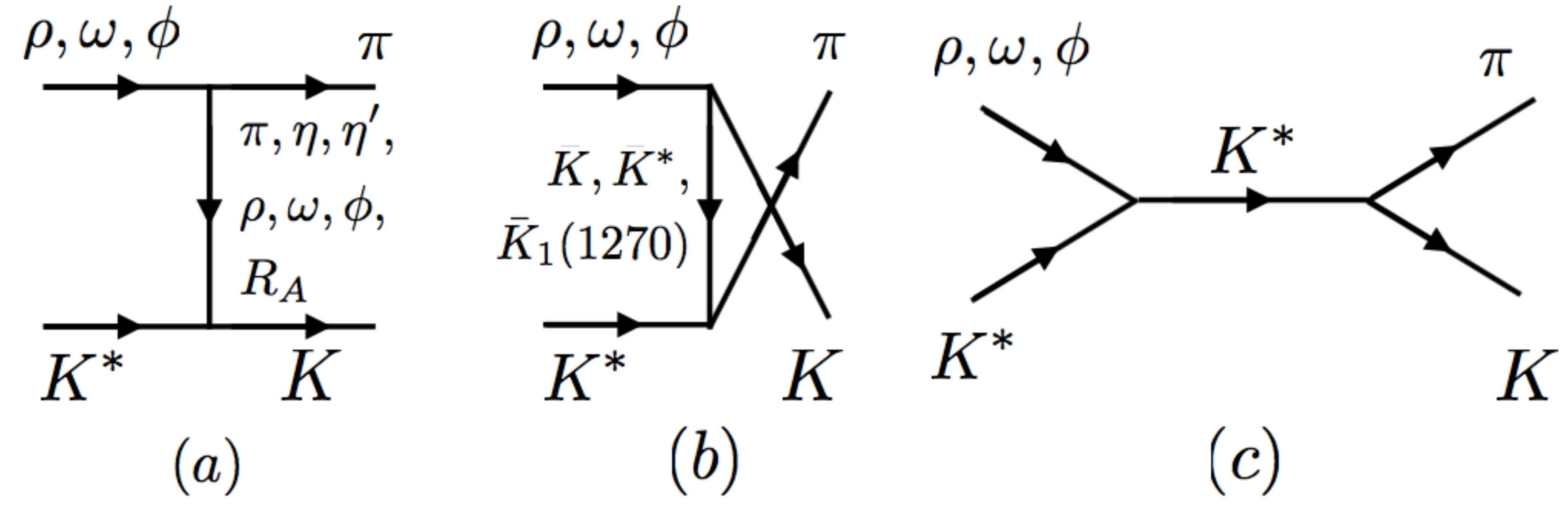}                        
\caption{Diagrams contributing to the processes 
$\rho K^*,\,\omega K^*,\, \phi K^*\to \pi K$ in the 
$t$-channel (a), $u$-channel (b) and $s$-channel (c). 
The symbol $R_A$ represents the exchange of the axial resonances 
$h_1(1170)$, $h_1(1380)$, $f_1(1285)$, $a_1(1260)$ and $b_1(1235)$ 
listed by the PDG and which are found from the dynamics in the 
$\bar K^* K$ system and coupled channels~\cite{rocaS}.}\label{rhoKstar}
\end{figure}

In Figs.~\ref{piKstar} and~\ref{rhoKstar} we show the different diagrams 
contributing to the processes $\pi K^*\to \rho K, \omega K,\phi K$ and 
$\rho K^*,\omega K^*, \phi K^*\to \pi K$ (without specifying the charge 
of the reaction).
Each of the amplitudes $\mathcal{M}^{ij}$ of Eq.~(\ref{Mqq}) can be written as
\begin{align}
\mathcal{M}^{ij}=T^{ij}+U^{ij}+S^{ij},
\end{align}
where $T^{ij}$, $U^{ij}$ and $S^{ij}$ are the contributions related to the 
$t$-, $u$- and $s$- channel diagrams shown in Figs.~\ref{piKstar} 
and~\ref{rhoKstar} for the process $i\to j$ for a particular total charge of 
the reaction $r$.

The amplitudes for these $t$-, $u$- and $s$- channel diagrams are determined 
by considering Lagrangians for the Pseudoscalar-Pseudoscalar-Vector (PPV), 
Vector-Vector-Pseudoscalar  (VVP) and Vector-Vector-Vector (VVV) vertices. 
These Lagrangians are based on an effective theory in which the vector mesons 
are identified as the dynamical gauge bosons of the hidden $\textrm{U}(3)_V$  
local symmetry in the $\textrm{U}(3)_L\times \textrm{U}(3)_R/\textrm{U}(3)_V$ 
non-linear sigma model~\cite{Bando1,Bando2,Meissner,Harada}, obtaining
\begin{align}
\mathcal{L}_{PPV}&=-i g_{PPV}  \langle V^\mu [P,\partial_\mu P]\rangle,\nonumber\\
\mathcal{L}_{VVP}&=\frac{g_{VVP}}{\sqrt{2}}\epsilon^{\mu\nu\alpha\beta}
\langle\partial_\mu V_\nu\partial_\alpha V_\beta P\rangle\label{Lag}\\
\mathcal{L}_{VVV}&=i g_{VVV} \langle(V^\mu\partial_\nu V_\mu-
\partial_\nu V_\mu V^\mu) V^\nu)\rangle.\nonumber
\end{align} 
The $\mathcal{L}_{VVP}$ Lagrangian written above contains the Levi-Civita 
pseudotensor since it describes an anomalous vertex, which involves a violation 
of the natural parity in the vertex~\cite{wess,witten}. In Eq.~(\ref{Lag}), $P$ 
and $V_\mu$ are matrices containing the octet of pseudoscalars and vectors mesons 
and the singlet of SU(3), respectively, which in the physical basis and considering 
ideal mixing for $\eta$ and $\eta^\prime$ as well as for $\omega$ and $\phi$ read 
as~\cite{rocaS, gengR, Daniel}:
\begin{align}
P&=\left(\begin{array}{ccc}
\frac{\eta}{\sqrt{3}}+\frac{\eta^\prime}{\sqrt{6}}+\frac{\pi^0}{\sqrt{2}}&\pi^+&K^+\\ 
\pi^- & \frac{\eta}{\sqrt{3}}+\frac{\eta^\prime}{\sqrt{6}}-\frac{\pi^0}{\sqrt{2}}&K^0\\
K^-&\bar{K}^0&-\frac{\eta}{\sqrt{3}}+\sqrt{\frac{2}{3}}\eta^\prime
\end{array}\right),
\end{align}
\begin{align}
V_\mu&=\left(\begin{array}{ccc}
\frac{\omega+\rho^0}{\sqrt{2}}&\rho^+&K^{*+}\\ 
\rho^- &\frac{\omega-\rho^0}{\sqrt{2}}&K^{*0}\\
K^{*-}&\bar{K}^{*0}&\phi
\end{array}\right)_\mu.
\end{align}

The couplings appearing in Eq.~(\ref{Lag}) are given by~\cite{Nagahiro1,Khem,Khem2}
\begin{align}
g_{PPV}&=\frac{m_V}{2 f_\pi},\quad
g_{VVP}=\frac{3 m^2_V}{16 \pi^2 f^3_\pi},\quad
g_{VVV}=\frac{m_V}{2 f_\pi},\label{coup}
\end{align}
with $m_V$ being the mass of the vector meson, which we take as the mass of the 
$\rho$ meson, and $f_\pi=93$ MeV is the pion decay constant. 
The symbol $\langle\,\rangle$  in Eq.~(\ref{Lag}) indicates the trace in the 
isospin space. 

The evaluation of some of the diagrams in Figs.~\ref{piKstar} and~\ref{rhoKstar} 
requires the coupling of several axial resonances to their hadron components. This 
couplings are directly taken from Refs.~\cite{rocaS, gengR}, and we list them in 
Tables~\ref{K1} and \ref{RA} of the Appendix~\ref{apenA} for the convenience of 
the reader.

After defining all the ingredients needed for the evaluation of the contribution 
of the diagrams in Figs.~\ref{piKstar} and \ref{rhoKstar}, we can start writing 
the contributions explicitly. The $t$-channel, $T^{ij}$, $u$-channel, $U^{ij}$, 
and $s$-channel, $S^{ij}$, amplitudes for the diagrams shown in               
Figs.~\ref{piKstar}a,~\ref{piKstar}b, and ~\ref{piKstar}c, respectively, for a 
reaction $r$ of the type $i\,(a+b)\to j\,(c+d)$ are given by
\begin{align}
T^{ij}&=\sum_k \mathbb{T}^{\,ij}_kg^2_{PPV}\epsilon^\mu(k)\epsilon^\nu(p^\prime) 
k^\prime_\mu p_\nu\frac{1}{t-m^2_{P_k}+i\epsilon}\nonumber\\
&\quad+\sum_k \mathbb{\overline{T}}^{\,ij}_k\,g^2_{VVP}\epsilon^{\mu\nu\alpha\beta}
\epsilon^{\mu^\prime\nu^\prime\alpha^\prime}_{\phantom{\mu^\prime\nu^\prime\alpha^\prime}\beta} 
\,p^\prime_\mu p_\alpha k_{\mu^\prime}k^\prime_{\alpha^\prime}\epsilon_\nu(p^\prime)
\epsilon_{\nu^\prime}(k)\frac{1}{t-m^2_{V_k}+i\,\Gamma_{V_k} m_{V_k}},\label{Tpi}
\end{align}
\begin{align}
U^{ij}&=\sum_k \mathbb{U}^{ij}_kg_{VVV}g_{PPV}\frac{1}{u-m^2_{V_k}+i\,
\Gamma_{V_k}m_{V_k}}\epsilon^\mu(k)\epsilon^\nu(p^\prime)\nonumber\\
&\quad\times\left[2\left(-1+\frac{m^2_\pi-m^2_{K}}{m^2_{V_k}}\right)
(k_\nu p_\mu+p^\prime_\mu p_\nu)-2\left(1+\frac{m^2_\pi-m^2_K}{m^2_{V_k}}\right)
(k_\nu k^\prime_\mu+p^\prime_\mu k^\prime_\nu)\right.\nonumber\\
&\quad\quad-\left.g_{\nu\mu}\left\{-2(k^\prime+p)\cdot p^\prime+(m^2_\pi-m^2_K)
\left(1+\frac{m^2_\rho-m^2_{K^*}}{m^2_{V_k}}\right)\right\}\right],\label{Upi}
\end{align}
\begin{align}
S^{ij}&= \mathbb{S}^{ij}_{K}g^2_{PPV}\epsilon^\mu(k)\epsilon^\nu(p^\prime)p_\mu 
k^\prime_\nu \frac{1}{s-m^2_K+i\epsilon}\nonumber\\
&\quad+\mathbb{S}^{ij}_{K^*}g^2_{VVP}\epsilon^{\mu\nu\alpha\beta}
\epsilon^{\mu^\prime\nu^\prime\alpha^\prime}_
{\phantom{\mu^\prime\nu^\prime\alpha^\prime}\beta}\epsilon_\nu(k)
\epsilon_{\nu^\prime}(p^\prime)k_\mu 
p^\prime_{\mu^\prime}p_\alpha k^\prime_{\alpha^\prime}\frac{1}{s-m^2_{K^*}
+i\,\Gamma_{K^*}m_{K^*}}+\nonumber\\
&\quad+\sum_{l=1}^2 \frac{g^{(i)}_{K_1,l}g^{(j)}_{K_1,l}}{s-M^2_{K_1,l}+i\,\Gamma_{K_1,l}M_{K_1,l}}
\left[-g_{\mu\nu}+\frac{p_\mu k^\prime_\nu}{M^2_{K_1,l}}\right]
\epsilon^\mu(k)\epsilon^\nu(p^\prime),\label{Spi}
\end{align}
where $\mathbb{T}^{\,ij}_k$, $\mathbb{\overline{T}}^{\,ij}_k$, 
$\mathbb{U}^{\,ij}_k$, $\mathbb{S}^{ij}_{K}$ and $\mathbb{S}^{ij}_{K^*}$ are 
coefficients which depend on the initial $i$ and final $j$ channels, as well as 
the exchanged particle $k$, and they are given in Tables~\ref{CijT}-\ref{CijSKstar} 
of the Appendix~\ref{apenA}. In Eqs.~(\ref{Tpi}),~(\ref{Upi}) and ~(\ref{Spi}), 
$p$, $k$ are, respectively, the four-momentum of the $\pi$ and $K^*$ in the initial 
state, and $p^\prime$ and $k^\prime$ correspond, respectively, to the four-momentum  
of the vector meson ($\rho$, $\omega$ or $\phi$) and the $K$ in the final state, 
$\epsilon^{\mu\nu\alpha\beta}$ is the Levi-Civita tensor and $\epsilon_\mu(q)$ is 
the polarization vector associated with the particle exchanged, with four momentum 
$q$. To arrive to these expressions we have made use of the Lorenz gauge, in which 
$\epsilon(p)\cdot p=0$, and the fact that the contraction of an antisymmetric 
tensor, like the Levi-Civita tensor, with a symmetric one gives 0. 
The $m_{P_k}$ in Eq.~(\ref{Tpi}) corresponds to the mass of the exchanged         
pseudoscalar in Fig.~\ref{piKstar}a and $m_{V_k}$ and $\Gamma_{V_k}$ are the mass 
and width, respectively, of the exchanged vector. We have considered \cite{PDG}:  
$\Gamma_\phi=4.3$ MeV, $\Gamma_\omega=8.5$ MeV, $\Gamma_{K^*}=50.5$ MeV and 
$\Gamma_\rho=149.4$ MeV and used isospin average masses, $m_\rho=770$ MeV, 
$m_\omega=782$ MeV, $m_{K^*}=892$ MeV, $m_\phi=1020$ MeV, $m_\pi=137$ MeV and $m_K=496$ MeV. 

In Eqs.~(\ref{Upi}) and (\ref{Spi}), $g_{\alpha\beta}$ is the Minkowski metric 
tensor. The $M_{K_1,l}$, $\Gamma_{K_1,l}$ and $g^{(i)}_{K_1,l}$ and 
$g^{(j)}_{K_1,l}$ present in Eq.~(\ref{Spi}) are the mass, width and coupling 
of the pole $l$ (to the initial $i$ and final $j$ channels) associated with the 
axial state $K_1(1270)$. These values can be found in Table~\ref{K1} of the 
Appendix~\ref{apenA}. In case of the $t$-channel amplitude of Eq.~(\ref{Tpi}), 
we have considered the exchange of pseudoscalars as well as vector mesons. 
A note here is in order. When exchanging a pion in the $t$-channel in the 
reaction $\pi K^*\to \rho K$, the energy-momentum conservation in the vertex 
$\pi\to \pi\rho$ of Fig.~\ref{piKstar}a is such that the exchanged pion can 
become on-shell. Because of this, in some regions of the phase-space,  the  
pion propagator develops a pole originating a singular cross section
~\cite{Peierls,Chung}. This latter singularity in the cross section can be 
removed by the so-called Peierls method~\cite{Peierls}, where the basic idea 
is to introduce a complex four-momentum for the unstable particle in the vertex 
by considering its decay width. As a consequence, the four-momentum of the 
exchanged particle gets an imaginary part through the energy-momentum conservation, 
which leads to~\cite{Peierls,Chung}
\begin{align}
\frac{1}{t-m^2_\pi+i\epsilon}\to\frac{1}{t-m^2_\pi-im_\rho \Gamma_\rho\frac{E_\rho-E_\pi}{E_\rho}},
\end{align}
where $E_\pi$ and $E_\rho$ are the energies for the external $\rho$ and $\pi$ in 
the center of mass frame. For the $u$-channel amplitude we can only have exchange 
of vector mesons, since the exchange of a pseudoscalar meson implies a vertex which 
would violate either parity or angular momentum. For the case of the $s$-channel 
amplitude we have considered exchange of pseudoscalars, vector mesons and 
resonances, with the only possibilities compatible with conservation of quantum 
numbers being the pseudoscalar $K$, the vector $K^*$ and the state $K_1(1270)$.

In the case of the  $t$-, $u$- and $s-$ channel diagrams in 
Figs.~\ref{rhoKstar}a$-$\ref{rhoKstar}c, respectively, we find the following 
contributions
\begin{align}
\pmb{T}^{ij}&=\sum_k\mathcal{T}^{ij}_kg^2_{PPV}\epsilon^\mu(p)\epsilon^\nu(k)p^\prime_\mu 
k^\prime_\nu\frac{1}{t-m^2_{P_k}+i\epsilon}\nonumber\\
&\quad+\sum_k\mathcal{\overline{T}}^{ij}_kg^2_{VVP}\epsilon^{\mu\nu\alpha\beta}
\epsilon^{\mu^\prime\nu^\prime\alpha^\prime}_{\phantom{\mu^\prime\nu^\prime\alpha^\prime}\beta}
\epsilon_\nu(p)\epsilon_{\nu^\prime}(k)p_\mu p^\prime_\alpha k_{\mu^\prime}
k^\prime_{\alpha^\prime}\frac{1}{t-m^2_{V_k}
+i\,\Gamma_{V_k} m_{V_k}}\nonumber\\
&\quad-\sum_{A}\frac{g^{(1)}_{A} g^{(2)}_{A}}{t-M^2_{A}+i\,\Gamma_{A}M_{A}}
\left[g_{\mu\nu}+\frac{p^\prime_\mu k^\prime_\nu}{M^2_{A}}\right]\epsilon^\mu(p)\epsilon^\nu(k),\label{Trho}
\end{align}
\begin{align}
\pmb{U}^{ij}&=\mathcal{U}^{ij}_{\bar K}g^2_{PPV}\epsilon^\mu(p)\epsilon^\nu(k)k^\prime_\mu 
p^\prime_\nu\frac{1}{u-m^2_{\bar K}+i\epsilon}\nonumber\\
&\quad+\mathcal{U}^{ij}_{\bar K^*}g^2_{VVP}\epsilon^{\mu\nu\alpha\beta}
\epsilon^{\mu^\prime\nu^\prime\alpha^\prime}_{\phantom{\mu^\prime\nu^\prime\alpha^\prime}\beta}
\epsilon_\nu(p)\epsilon_{\nu^\prime}(k)p_\mu k^\prime_\alpha k_{\mu^\prime}p^\prime_{\alpha^\prime}
\frac{1}{u-m^2_{\bar K^*}+i\Gamma_{\bar K^*}m_{\bar K^*}}\nonumber\\
&\quad-\sum_{l=1}^2\frac{g^{(1)}_{K_1,l}g^{(2)}_{K_1,l}}{u-M^2_{K_1,l}+i\,\Gamma_{K_1,l}M_{K_1,l}}
\left[g_{\mu\nu}+\frac{k^\prime_\mu p^\prime_\nu}{M^2_{K_1,l}}\right]\epsilon^\mu(p)\epsilon^\nu(k),\label{Urho}
\end{align}
\begin{align}
\pmb{S}^{ij}&=\sum_k\mathcal{S}^{ij} g_{PPV}g_{VVV}\Bigg[\left\{(m^2_{K^*}-m^2_V)
\left(1-\frac{m^2_K-m^2_\pi}{m^2_{V_k}}\right)-2(k\cdot p^\prime-p\cdot p^\prime)\right\}
\epsilon(k)\cdot\epsilon(p)\nonumber\\
&\quad-2\left(k_\mu k^\prime_\nu+p^\prime_\mu p_\nu-k_\mu p^\prime_\nu -k^\prime_\mu p_\nu\right)\epsilon^\mu(p)
\epsilon^\nu(k)\Bigg]\frac{1}{s-m^2_{V_k}+i\,\Gamma_{V_k}m_{V_k}},\label{Srho}
\end{align}
where $\mathcal{T}^{ij}$, $\mathcal{\overline{T}}^{ij}$, 
$\mathcal{U}^{ij}_{\bar K}$, $\mathcal{U}^{ij}_{\bar K^*}$ and 
$\mathcal{S}^{ij}$ are coefficients which are given in 
Tables~\ref{CijTrho}$-$\ref{CijSrho} of the Appendix~\ref{apenA}. 
In Eqs.~(\ref{Trho}),~(\ref{Urho}),~(\ref{Srho}), and all diagrams 
depicted in Fig.~\ref{rhoKstar}, $p$, $k$, are, respectively, the four 
momenta of the external vector meson without strangeness 
($\rho$, $\omega$, or $\phi$) and of the external $K^*$, while $p^\prime$ and 
$k^\prime$ are the four momenta of the external $\pi$ and $K$, respectively. 
The symbols $M_A$, $\Gamma_A$, $g^{(1)}_A$ and $g^{(2)}_A$ in Eq.~(\ref{Trho}) 
represent, respectively, the mass, width and coupling constants to the two 
vertices shown in Fig.~\ref{rhoKstar}a for the pole associated with the exchanged 
axial resonance $R_A$ (see Table~\ref{RA} of the Appendix~\ref{apenA} for the 
numerical values). In Eq.~(\ref{Urho}), $M_{K_1,l}$, $\Gamma_{K_1,l}$, 
$g^{(1)}_{K_1,l}$ and $g^{(2)}_{K_1,l}$ correspond to the mass, width and 
coupling constants to the two vertices shown in Fig.~\ref{rhoKstar}b for the 
pole $l$ related to the $K_1(1270)$ state and their numerical values are 
listed in Table~\ref{K1} of the Appendix~\ref{apenA}. In Eq.~(\ref{Srho}), 
$m_V$ is the mass of the external $\rho$, $\omega$ or $\phi$ vector mesons 
and, as in case of Eq.~(\ref{Upi}), $m_{V_k}$ and $\Gamma_{V_k}$ are the mass 
and width, respectively, of the exchanged vector meson in the diagram of 
Fig.~\ref{rhoKstar}b. Note that in this case we can not have the exchange of 
a pseudoscalar meson in the $s$-channel, since it implies the presence of a 
three pseudoscalar vertex.

\section{Results}

\subsection{Cross sections for the processes 
$\pi K^*\to \rho K,\, \omega K,\, \phi K$.}

\begin{figure}[h!]
\centering
\includegraphics[width=0.5\textwidth]{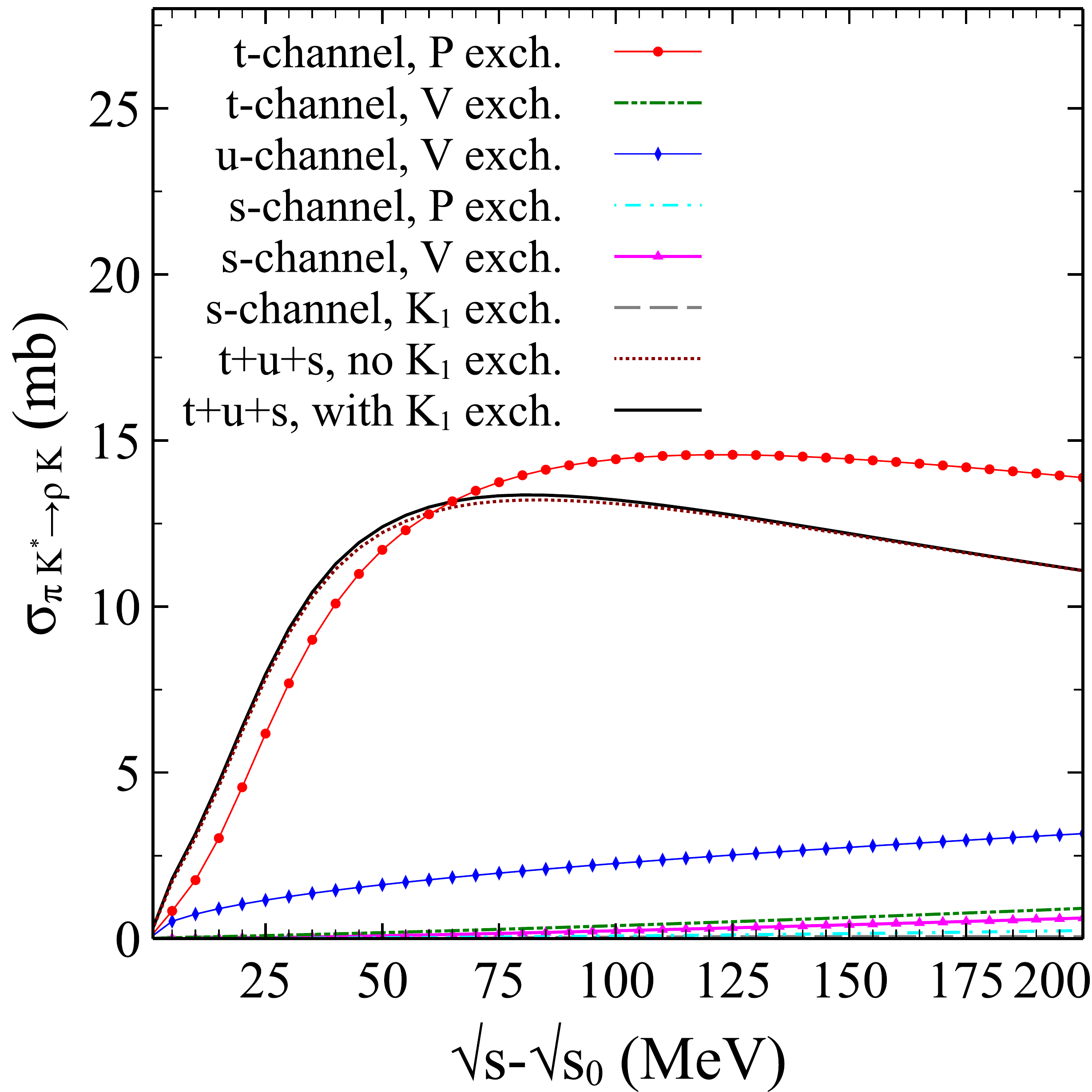}              
\caption{Cross sections obtained with Eq.~(\ref{cross}) for the process 
$\pi K^*\to \rho K$ considering different mechanism as a function of 
$\sqrt{s}-\sqrt{s_0}$, where $\sqrt{s}$ is the center of mass energy 
and $\sqrt{s_0}$ is the threshold energy for the reaction. }
\label{sigpiKstartorhoK}
\end{figure}

We start the discussion of the results by showing, in Fig.~\ref{sigpiKstartorhoK}, 
the cross sections obtained from Eq.~(\ref{cross}) for the reaction 
$\pi K^*\to \rho K$ with different mechanisms: (1) $t$-channel exchange of a 
pseudoscalar meson (solid line with circles); (2) $t$-channel exchange of a 
vector meson (dashed-dotted-dotted line); (3) $u$-channel exchange of a 
vector (line with rhombus); (4) $s$-channel exchange of a pseudoscalar 
(dashed-dotted line); (5) $s$-channel exchange of a vector (solid line 
with triangles) (6) $s$-channel exchange of $K_1(1270)$ (dashed line). 
The dotted and solid lines in Fig.~\ref{sigpiKstartorhoK} correspond,   
respectively, to considering the mechanisms (1)-(5) and (1)-(6) for the 
determination of the cross section. 

As can be seen in Fig.~\ref{sigpiKstartorhoK}, the contribution from the 
$t$-channel exchange of a pseudoscalar meson (not considered in 
Ref.~\cite{cho} ) gives rise to the largest cross section and the other 
mechanisms considered produce small corrections to it. Note that due to 
a reordering of the particles in the vertices, the $t$-channel ($u$-channel) 
exchange in Ref.~\cite{cho} corresponds to the $u$-channel ($t$-channel) exchange 
in the present work to which we refer throughout the text. It is also 
interesting to notice that the $u$-channel exchange of a vector meson 
(considered in Ref.~\cite{cho}) leads to a larger cross section than 
that associated with the $t$-channel exchange of a vector meson (not 
evaluated in Ref.~\cite{cho}) and the $s$-channel exchange of a 
pseudoscalar. The process in which a vector meson is exchanged in the 
$s$-channel (not taken into account in Ref.~\cite{cho}) gives a larger  
contribution to the cross section when compared with the one arising from 
the exchange of a pseudoscalar in the $s$-channel (considered in 
Ref.~\cite{cho}). 
It should be mentioned that the contribution of the $K_1(1270)$ exchange in 
the $s$-channel to the cross section is negligible (compare the solid and 
dotted lines of Fig.~\ref{sigpiKstartorhoK}).

\begin{figure}[h!]
\centering
\includegraphics[width=0.495\textwidth]{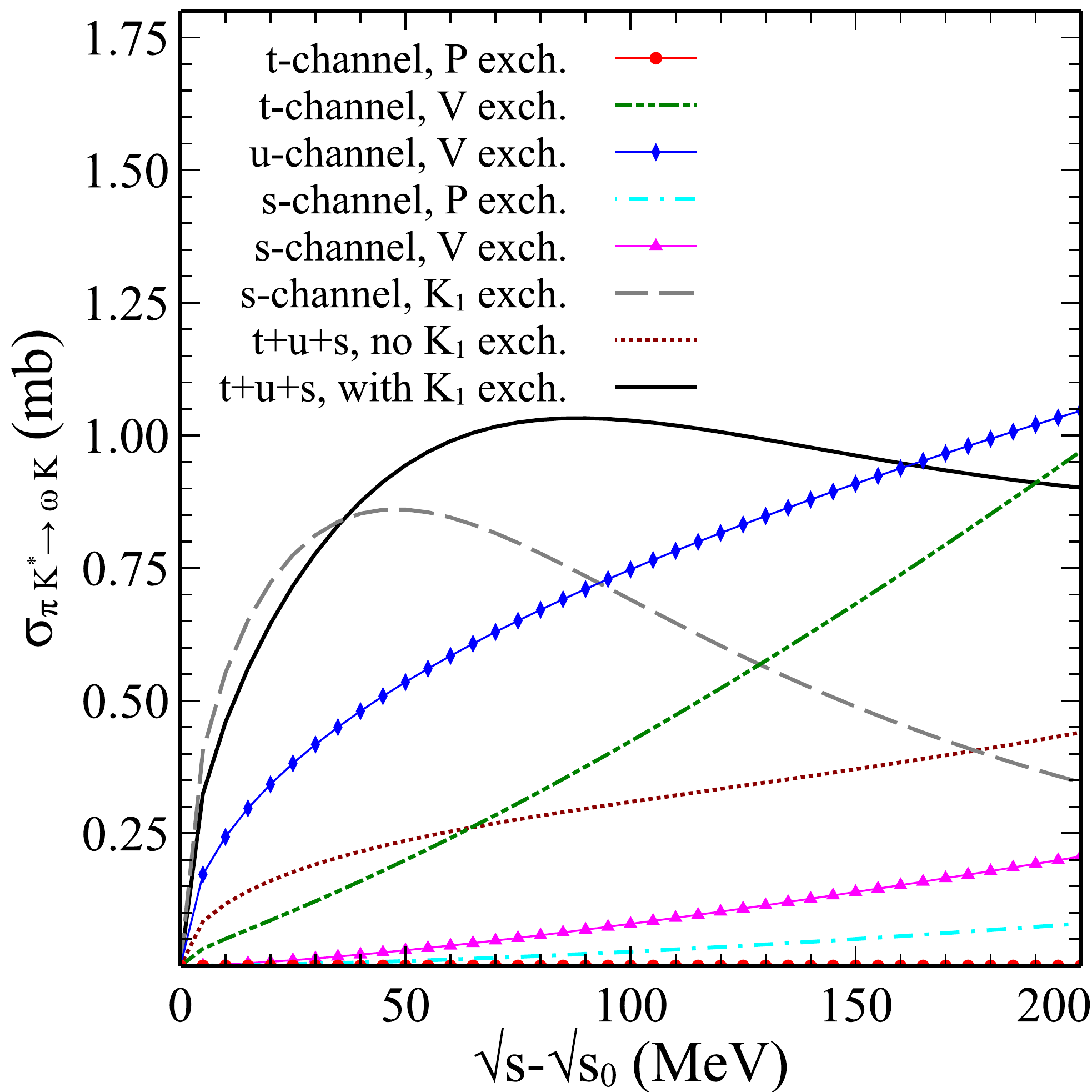}
\includegraphics[width=0.495\textwidth]{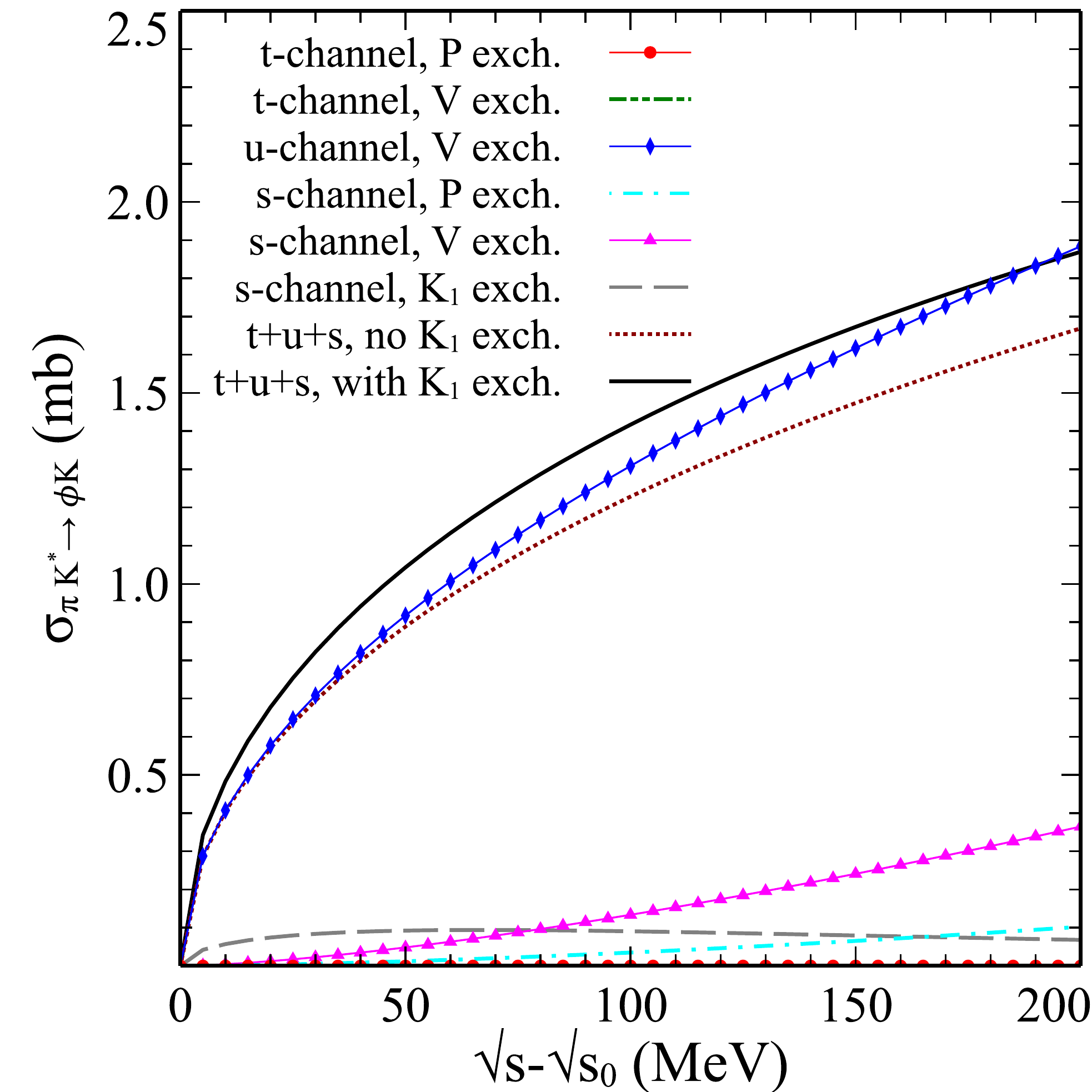}
\caption{Cross sections obtained with Eq.~(\ref{cross}) for the process 
$\pi K^*\to \omega K$ (left panel) and $\pi K^*\to\phi K$ (right panel) 
considering different mechanism as a function of $\sqrt{s}-\sqrt{s_0}$, 
where $\sqrt{s}$ is the center of mass energy and $\sqrt{s_0}$ is the 
threshold energy for the reaction.}\label{sigpiKstartowphiK}
\end{figure}

In Fig.~\ref{sigpiKstartowphiK}, 
we show the results for the reactions $\pi K^*\to \omega K$ (left panel) 
and $\pi K^*\to \phi K$ (right panel), reactions which were not considered 
in Ref.~\cite{cho}. As can be seen, the final cross section for both 
reactions (solid lines) have similar magnitude and both are smaller 
than the one for the process $\pi K^*\to\rho K$ (solid line in 
Fig.~\ref{sigpiKstartorhoK}) by around one order of magnitude. This finding 
indicates that the absorption 
mechanism of a $K^*$ by a pion, producing a $K$ together with an $\omega$ 
or a $\phi$  may probably not be relevant in the determination of the time 
evolution of the abundances found in Ref.~\cite{cho} for $K^*$ and $K$. 
Note, however, that without the contribution to the cross section of 
$\pi K^*\to\rho K$ from a diagram involving $\rho\pi\pi$ and $K^*\pi K$ 
vertices (not evaluated in Ref.~\cite{cho}), shown as line with circles 
in Fig.~\ref{sigpiKstartorhoK}, the cross sections for the processes 
$\pi K^*\to \rho K, \omega K, \phi K$ are comparable. It is also  
interesting to notice the relevance in the $\pi K^*\to \omega K$ cross 
section of the mechanism in which the $K_1(1270)$ is exchanged in the 
$s$-channel (dashed line in Fig.~\ref{sigpiKstartowphiK}, left panel). 
The inclusion of such $K_1(1270)$ exchange produces a significant change 
in the cross section, as can be noticed from Fig.~\ref{sigpiKstartowphiK} 
by comparing the dotted line, which shows the total cross section obtained  
without considering the exchange of $K_1(1270)$ in the $s$-channel, and the 
solid line, which corresponds to the result with such an exchange included.  
As in case of the process $\pi K^*\to \rho K$, the $u$-channel exchange of 
a vector meson in the reactions $\pi K^*\to\omega K$ and $\pi K^*\to\phi K$ 
(line with rhombus in both panels of Fig.~\ref{sigpiKstartowphiK}) gives larger 
contribution to the cross section than the exchange of a vector or pseudoscalar 
mesons in the $s$-channel or a vector meson in the $t$-channel, with the latter 
mechanism being more important in case of the process $\pi K^*\to\omega K$. 

\subsection{Cross sections for the processes $\rho K^*,\, \omega K^*,\, 
\phi K^*\to \pi K$.}

\begin{figure}[h!]
\centering
\includegraphics[width=0.55\textwidth]{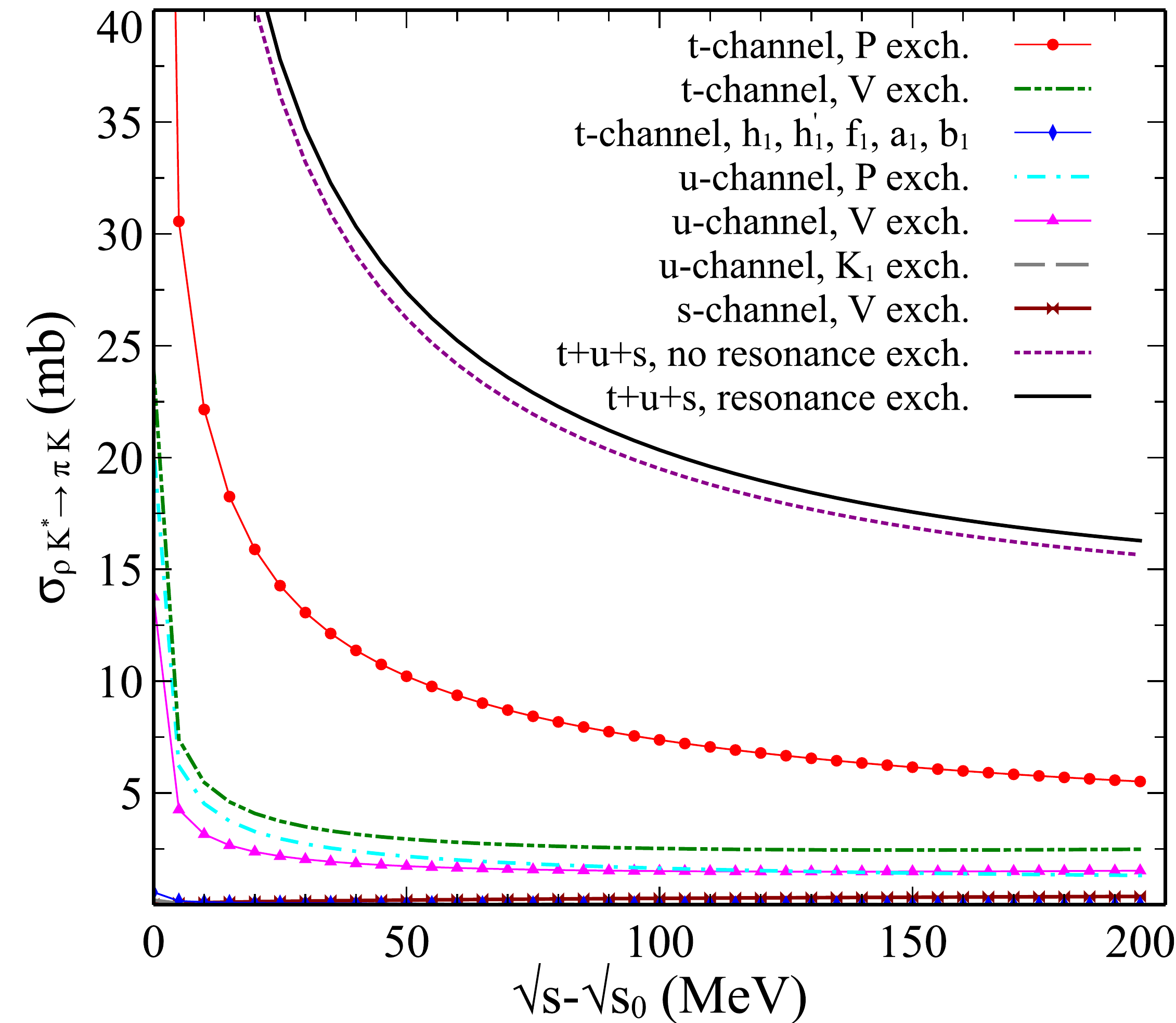}
\caption{Cross sections obtained with Eq.~(\ref{cross}) for the process 
$\rho K^*\to \pi K$ considering different mechanism as a function of $\sqrt{s}-\sqrt{s_0}$, 
where $\sqrt{s}$ is the center of mass energy and $\sqrt{s_0}$ is the threshold energy for the 
reaction. Since the cross section, independently of the mechanism, diverges at the threshold, 
they have been calculated at an energy starting 0.5 MeV above the threshold of the reaction, 
so the $x$-axis is actually $\sqrt{s}-\sqrt{s_0}+0.5\text{MeV}$. In the figure, $h_1\equiv h_1(1170)$ 
and $h^\prime_1\equiv h_1(1380)$.}\label{sigrhoKstar}
\end{figure}

In Fig.~\ref{sigrhoKstar} 
we show the cross section calculated with Eq.~(\ref{cross}) for the process $\rho K^*\to\pi K$ 
considering contributions from the mechanism shown in Figs.~\ref{rhoKstar}a$-$\ref{rhoKstar}c. Since 
the process is exothermic, the cross section diverges at the threshold. As can be seen, the consideration 
of the exchange of the resonances listed in Table~\ref{RA} in the $t$-channel and the exchange of $K_1(1270)$ 
in the $u$-channel produces a small modification in the total cross section. The contribution to the cross 
section from the exchange of a pseudoscalar meson in the $t$-channel is larger than that related to the 
exchange of a vector meson in the $t$- or $u$-channel (both missing in Ref.~\cite{cho}) and  that of a 
pseudoscalar meson in the $u$-channel (considered in Ref.~\cite{cho}). Since the $s$-channel exchange of a 
vector meson (taken into account in Ref.~\cite{cho}) turns out to give a very small contribution to the cross 
section, the other mechanisms considered here become  relevant.

\begin{figure}[h!]
\centering
\includegraphics[width=0.49\textwidth]{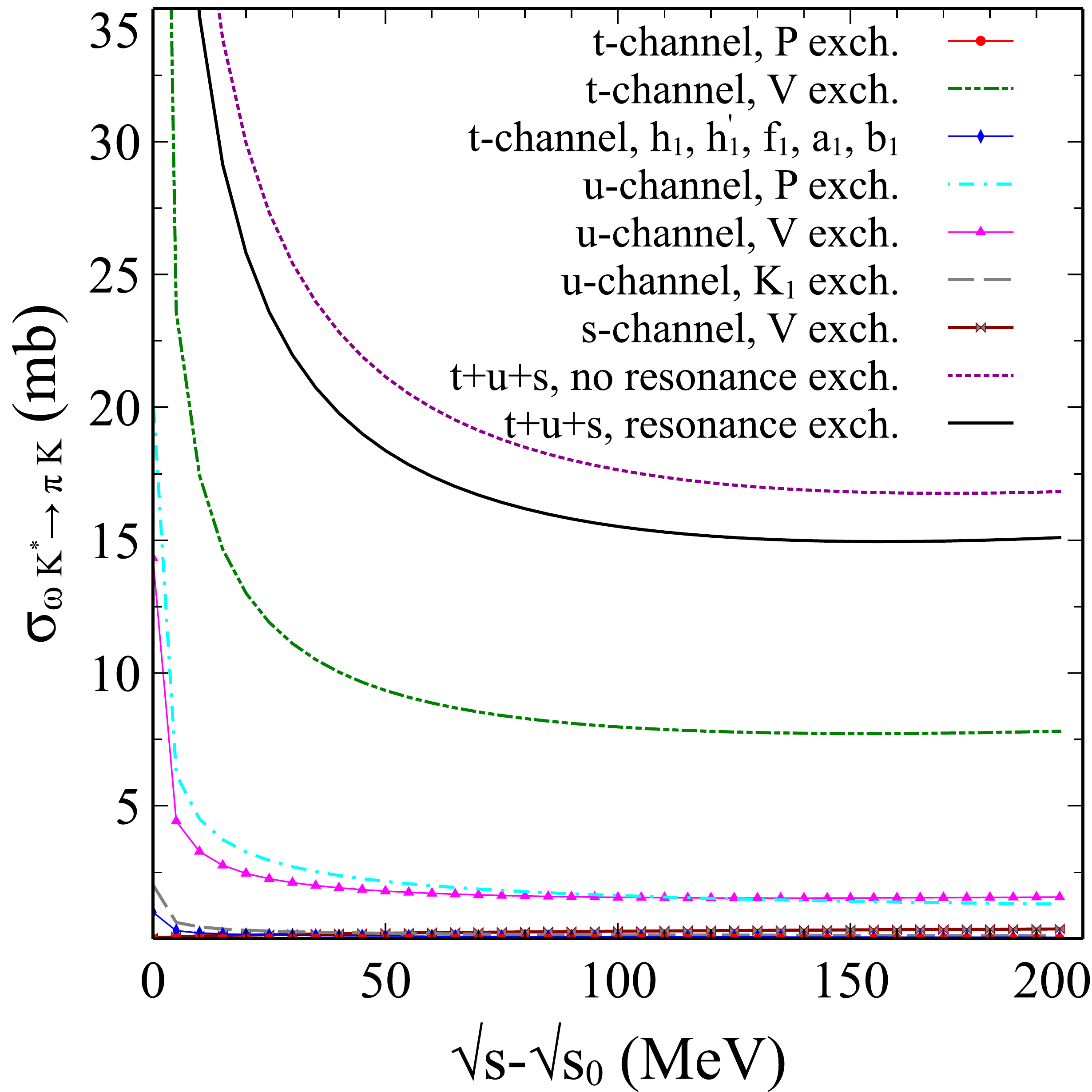}
\includegraphics[width=0.49\textwidth]{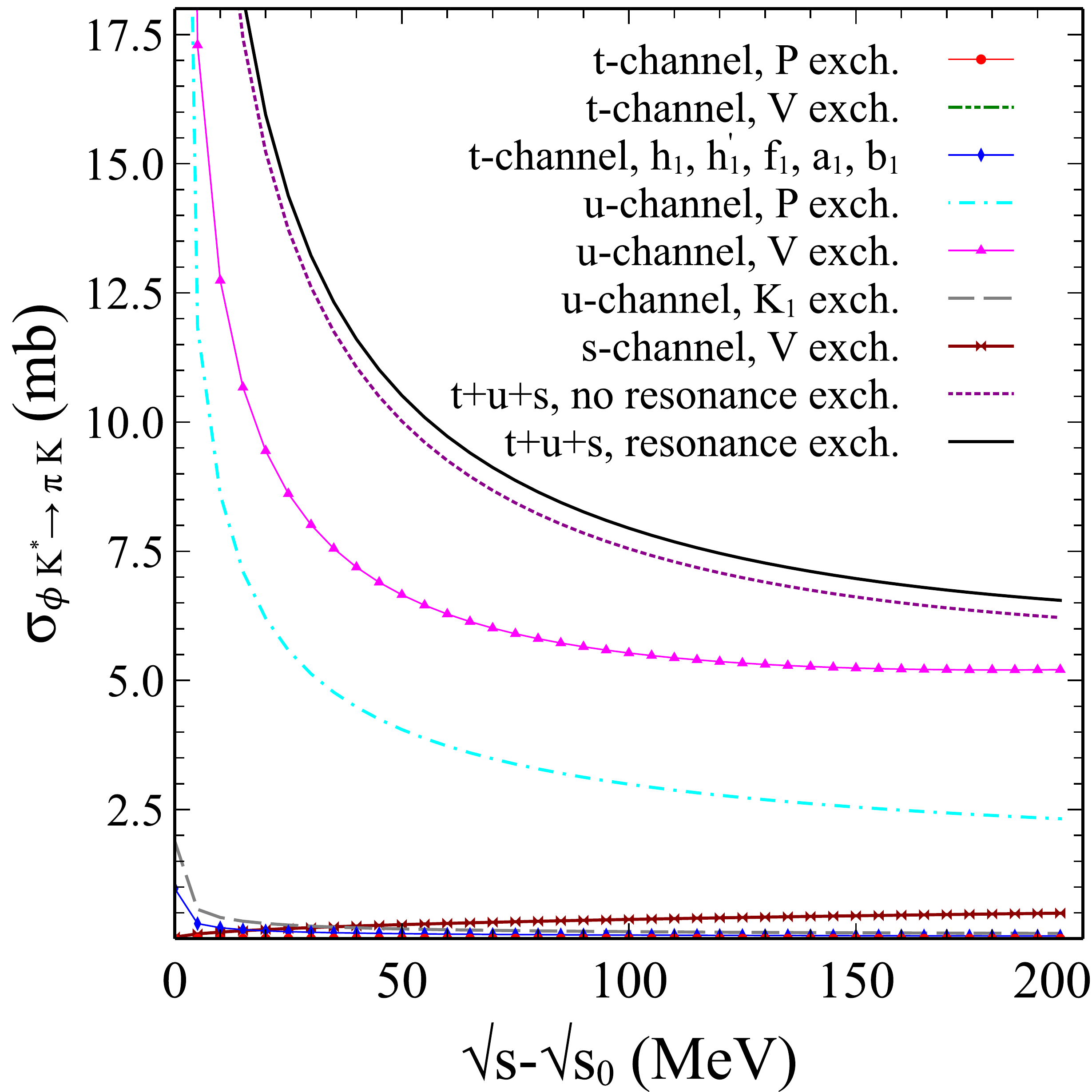}
\caption{Cross sections obtained with Eq.~(\ref{cross}) for the processes $\omega K^*\to \pi K$ (left panel) 
and $\phi K^*\to \pi K$ (right panel) considering different mechanism as a function of $\sqrt{s}-\sqrt{s_0}$, 
where $\sqrt{s}$ is the center of mass energy and $\sqrt{s_0}$ is the threshold energy for the reaction. Since 
the cross section, independently of the mechanism, diverges at threshold, they have been calculated at an energy 
starting 0.5 MeV above the threshold of the reaction, so the $x$-axis is actually $\sqrt{s}-\sqrt{s_0}+0.5\text{MeV}$. 
In the figure, $h_1\equiv h_1(1170)$ and $h^\prime_1\equiv h_1(1380)$.}\label{sigwphiKstar}
\end{figure}

Similar to the case $\rho K^*\to \pi K$, the resonance exchange in the $t$- and $u$- channels for the reactions 
$\omega K^*\to \pi K$ and $\phi K^*\to \pi K$ produces a weak modification in the cross section (compare the solid 
and dotted lines in both panels of Fig.~\ref{sigwphiKstar}). Interestingly, the final cross sections for 
$\rho K^*\to \pi K$, $\omega K^*\to \pi K$ and $\phi K^*\to\pi K$ have comparable magnitude. 

\subsection{Resonance exchange in $\rho K^*,\, \omega K^*,\, \phi K^*\to \pi K$ through triangular loops.}
\begin{figure}[h!]
\centering
\includegraphics[width=0.4\textwidth]{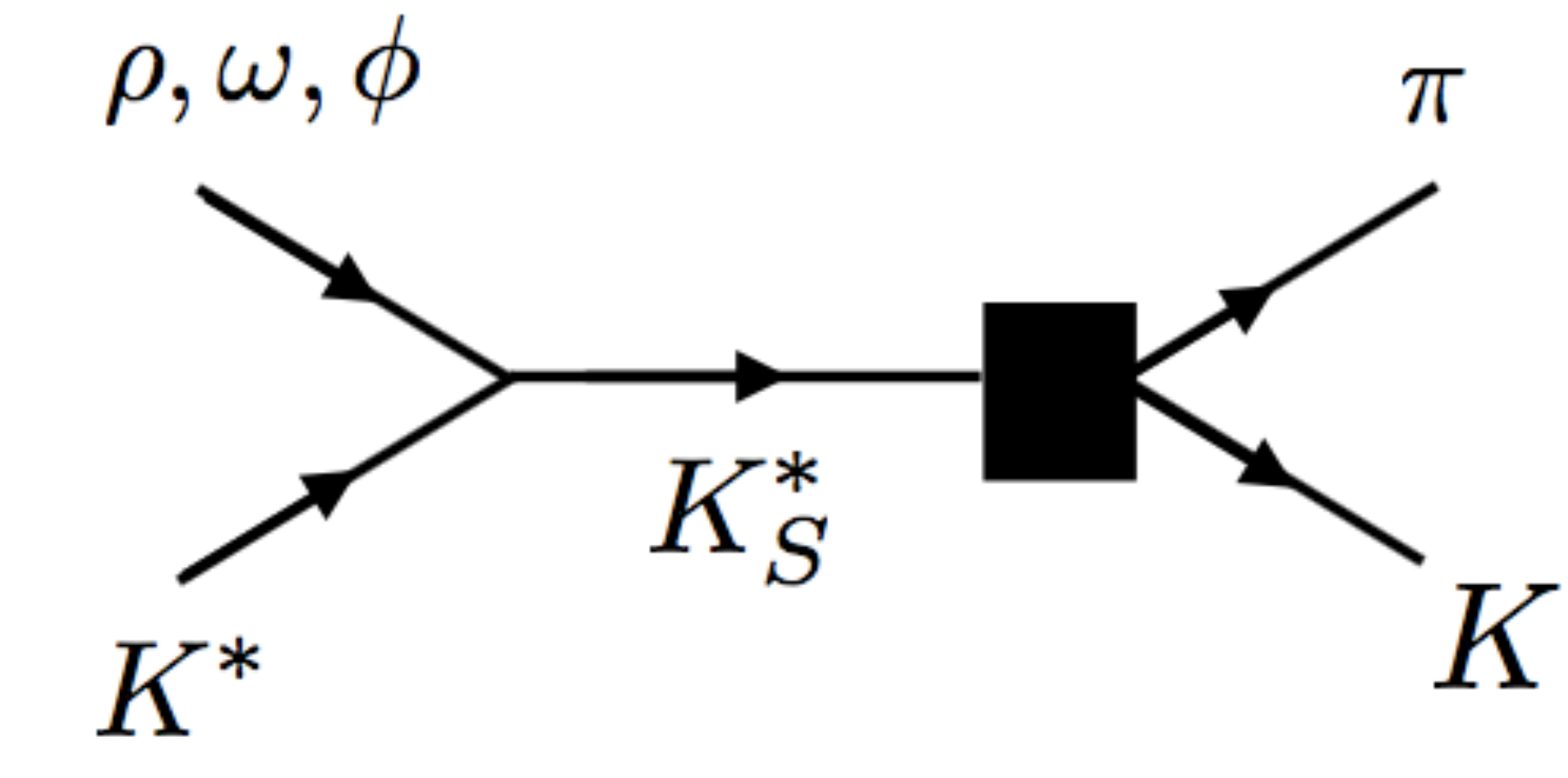}
\caption{Exchange of the $K^*_S$ states found in Ref.~\cite{gengo} in the $s$-channel for the reactions 
$\rho K^*,\,\omega K^*,\,\phi K^*\to \pi K$. The filled box in the figure corresponds to the vertex 
$K^*_S\pi K$, which is evaluated considering the triangular loop shown in Fig.~\ref{square}.}\label{res_sch_exch}
\end{figure}

\begin{figure}[h!]
\centering
\includegraphics[width=0.6\textwidth]{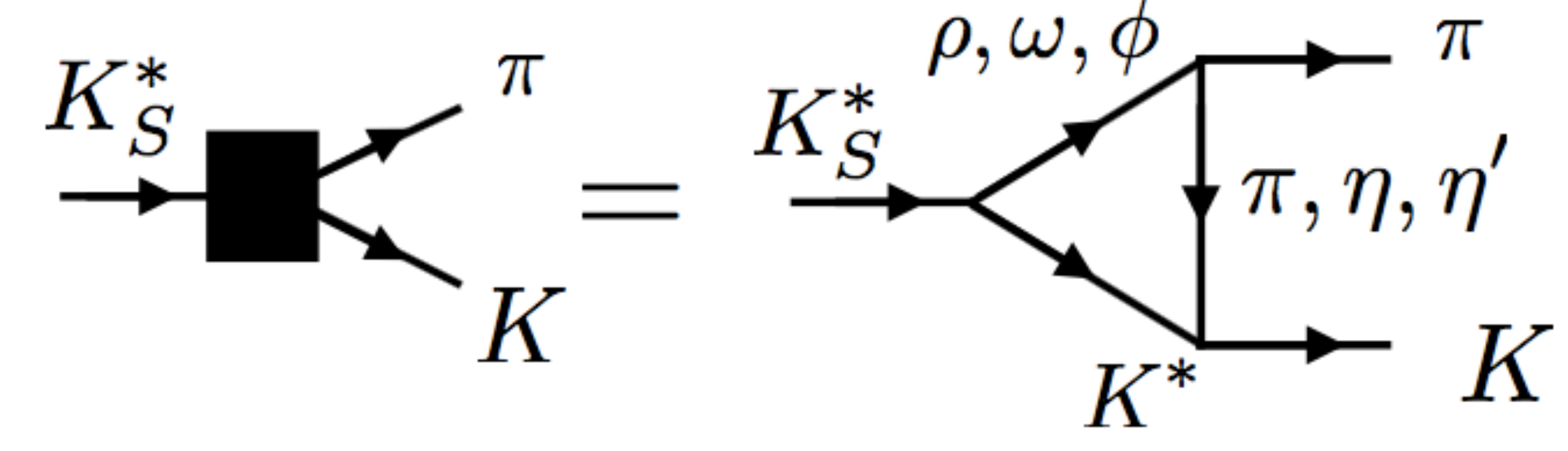}
\caption{The $K^*_S\pi K$ vertex.}\label{square}
\end{figure}

In addition to the mechanisms discussed so far to determine the cross sections of the reactions 
$\rho K^*,\, \omega K^*,\, \phi K^*\to \pi K$ (see Fig.~\ref{rhoKstar}), one could also consider the possibility of 
exchanging a resonance in the $s$-channel,
as in case of the $K_1(1270)$ exchange in $\pi K^*$ collisions (see Fig.~\ref{piKstar}).  
Indeed, in Ref.~\cite{gengo} the interaction of $K^*$ with $\rho$, $\omega$ and $\phi$ in $s$-wave 
(orbital angular momentum 0) was investigated and several $K^*$ resonances with $I=1/2$ and different spin were 
found as a consequence of the dynamics involved: a $J^P=0^+$ resonance with mass 1643 MeV and width of 48 MeV, which 
is a prediction of the theory; a $1^+$ resonance with mass 1737 MeV and width of 164 MeV which is associated with the 
state $K^*_1(1650)$ listed by the PDG~\cite{PDG}; a $J^P=2^+$ state with mass 1431 MeV and 56 MeV of width which is 
identified with the $K^*_2(1430)$ listed by the PDG. Thus, exchange of these $K^*_S$ states (with $S$ indicating the spin) 
in the $s$-channel, as shown in Fig.~\ref{res_sch_exch}, can be important while calculating the cross section for 
$\rho K^*\to \pi K$. As can be seen in Fig.~\ref{res_sch_exch}, one of the vertices involved in the process is the 
$K^*_S\pi K$ vertex. From Ref.~\cite{gengo}, we have information on the pole positions of these $K^*_S$ states and their 
couplings to the channels $\rho K^*$, $\omega K^*$ and $\phi K^*$ (which we list in Table~\ref{RV} of the Appendix~\ref{apenB}), 
but the couplings to two pseudoscalars are not available. However, one can still consider $K^*_S$ exchange in the $s$-channel 
through an effective vertex, represented by a filled box in Fig.~\ref{res_sch_exch}, by describing it through  triangular 
loops (see Fig.~\ref{square}).

The details related to the determination of the amplitude for the process depicted in Fig.~\ref{res_sch_exch} can be found in 
Appendix~\ref{apenB}. Since the interaction of the initial vector-vector system in the diagram of Fig.~\ref{res_sch_exch} 
would generate these $K^*_S$ states, the quantum numbers for the external vectors system can be $J^P=0^+$, $1^+$ or $2^+$. 
The final state in Fig.~\ref{res_sch_exch} consists of two pseudoscalars (total spin 0), thus, the only way of getting $J=1$ 
is with one unit of orbital angular momentum which leads the two pseudoscalar system to have $J^P=1^-$ instead of the 
initial $1^+$. This means that in the diagram of Fig.~\ref{res_sch_exch} we can not have a transition in the $s$-channel 
through the exchange of the $K^*_1$ resonance found in Ref.~\cite{gengo}. Similarly, we can not have interference between 
the diagrams in Fig.~\ref{rhoKstar}c, which involves the exchange of a pseudoscalar or vector meson (thus, a initial 
state having negative total parity), and the diagram in Fig.~\ref{res_sch_exch}.

\begin{figure}[h!]
\centering
\includegraphics[width=0.49\textwidth]{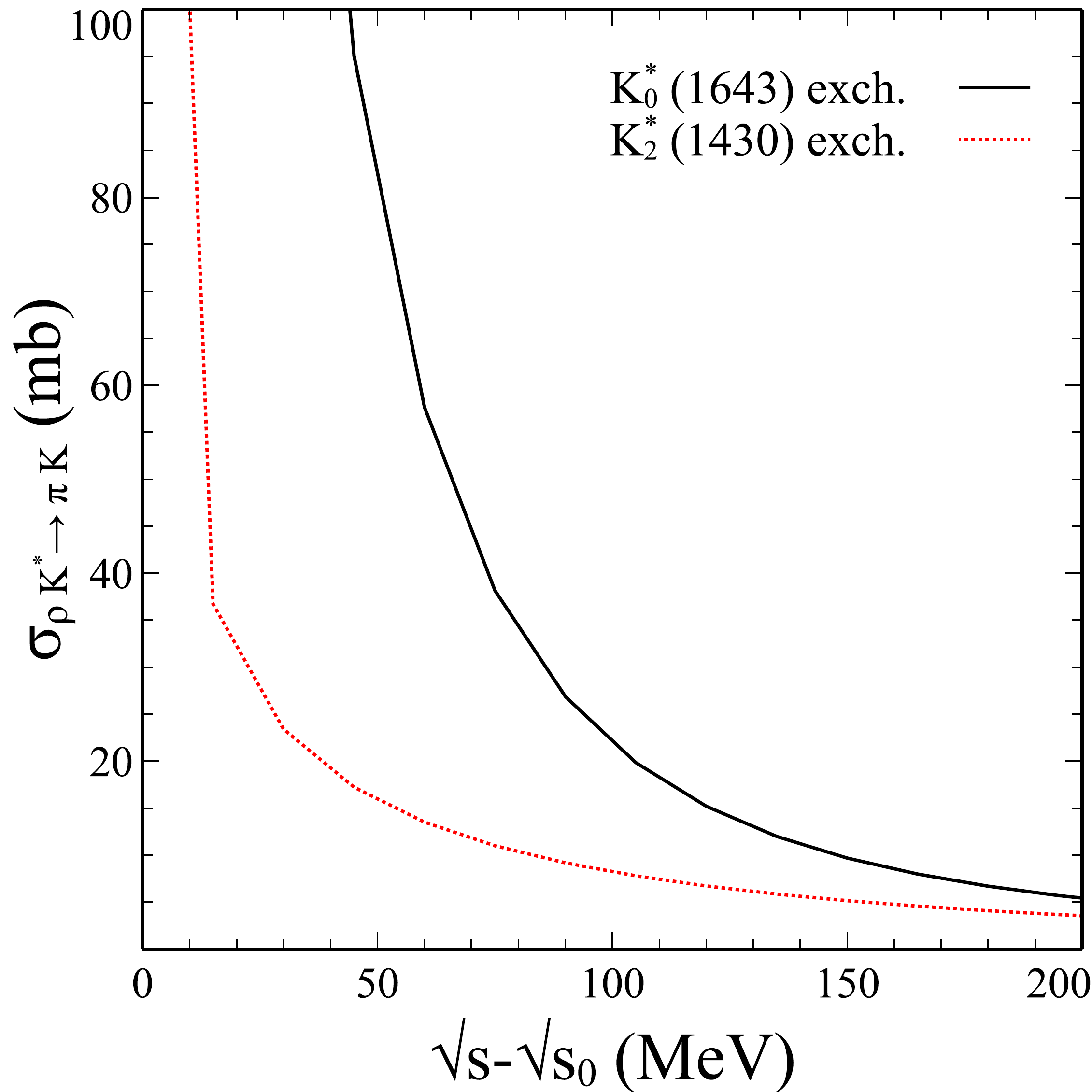}
\caption{Contribution to the cross section of $\rho K^*\to \pi K$ from the triangular loops shown in 
Fig.~\ref{square}, in which we consider the exchange of the states listed in Table~\ref{RV}. } 
\label{sigTrLoopsrhoKstar}
\end{figure}
\begin{figure}[h!]
\centering
\includegraphics[width=0.49\textwidth]{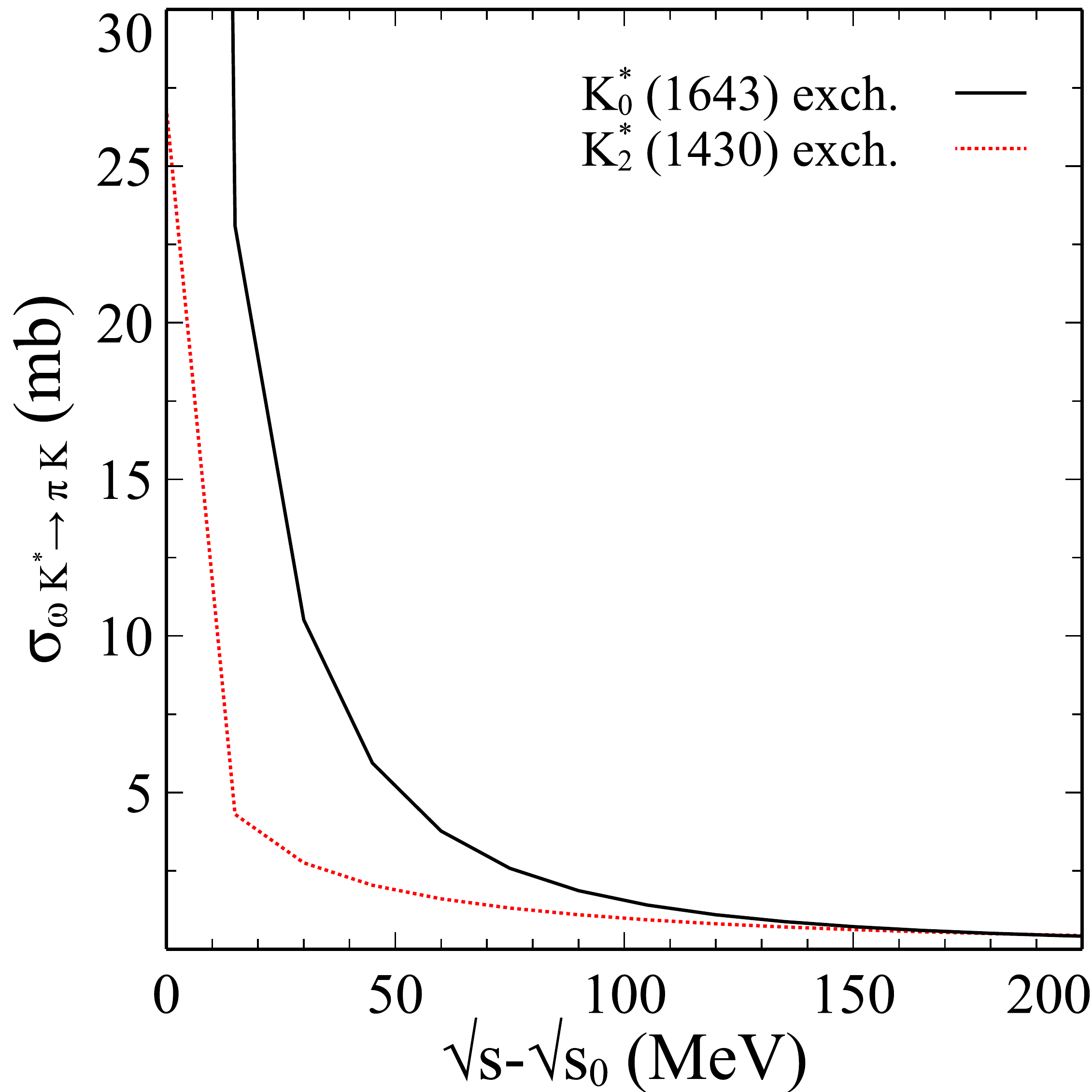}
\includegraphics[width=0.49\textwidth]{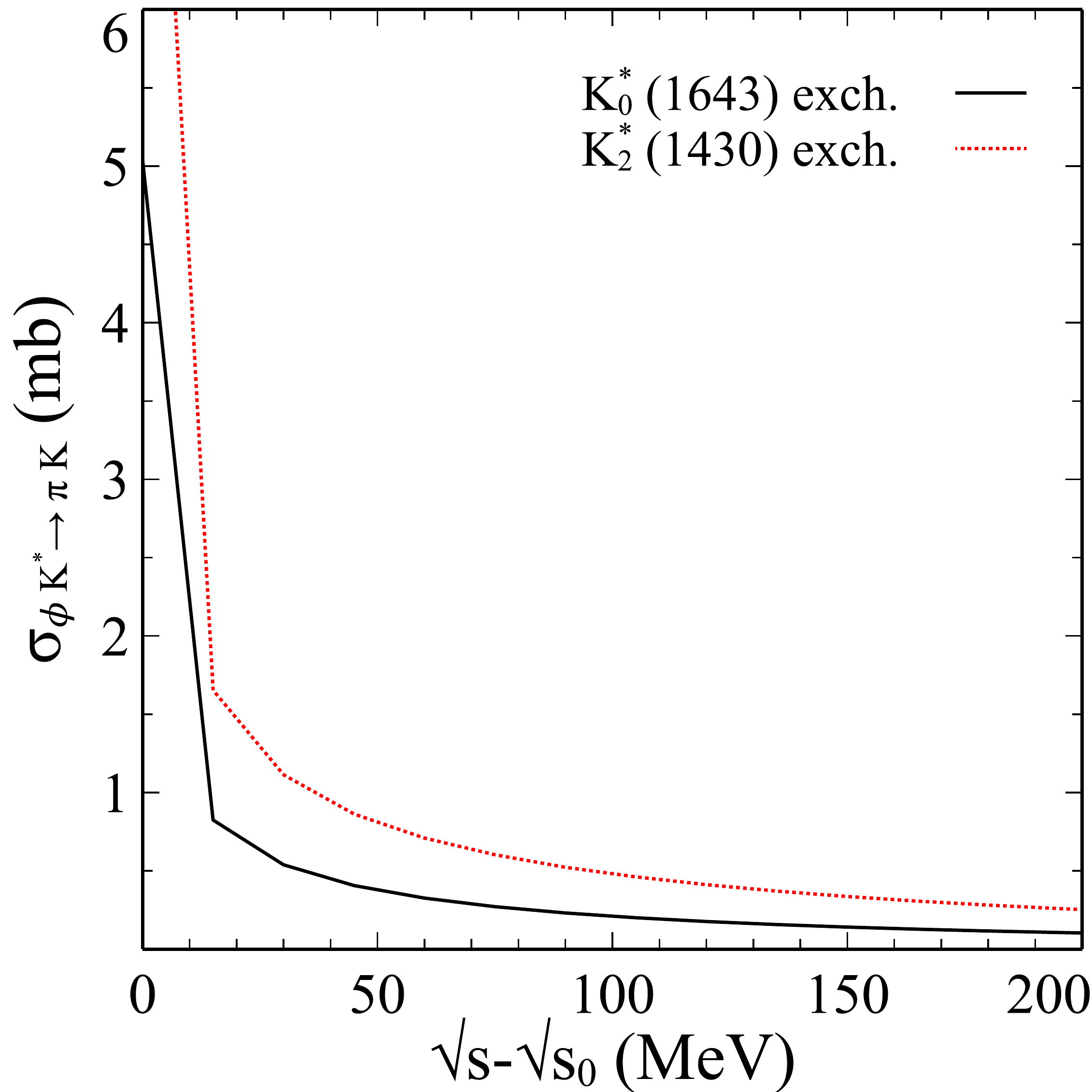}
\caption{Contribution to the cross section of $\omega K^*\to \pi K$ (left panel) and 
$\phi K^*\to\pi K$ (right panel) from the $s$-channel exchange of the $K^*_S$ states listed in Table~\ref{RV}.}\label{sigTrLoopswphiKstar}
\end{figure}
In Fig.~\ref{sigTrLoopsrhoKstar}
we show the cross section for the process $\rho K^*\to \pi K$ considering $s$-channel exchange of the $K^*_S$ resonances. 
As can be seen by comparing with the results shown in Fig.~\ref{sigrhoKstar}, the contribution to the cross section of the 
mechanism depicted in Fig.~\ref{res_sch_exch} is very relevant. These results suggest that the inclusion of the process 
shown in Fig.~\ref{res_sch_exch} must strongly affect the production of $K^*$ and $K$ in heavy ion collisions.

In Fig.~\ref{sigTrLoopswphiKstar} we show the results found for the cross section related to the $s$-channel exchange of $K^*_S$ 
in case of the reactions $\omega K^*\to \pi K$ (left panel) and $\phi K^*\to \pi K$ (right panel). By comparing with the results 
found in Fig.~\ref{sigwphiKstar}, this mechanism also produces changes in the cross section obtained without the $s$-channel 
$K^*_S$ exchange, although milder than in case of the reaction $\rho K^*\to \pi K$. 

\subsection{Cross sections for the inverse reactions}
\begin{figure}
\centering
\includegraphics[width=0.49\textwidth]{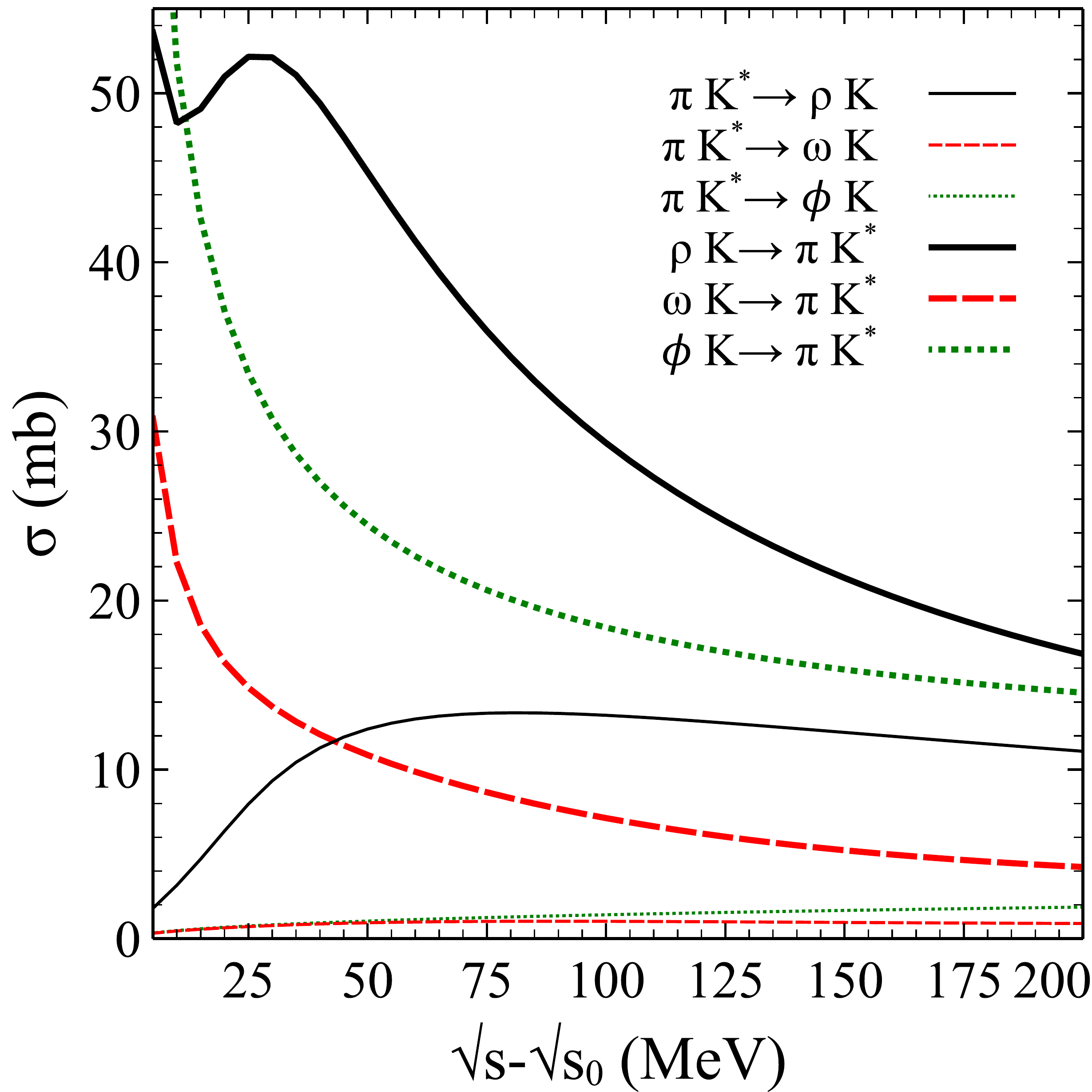}
\includegraphics[width=0.49\textwidth]{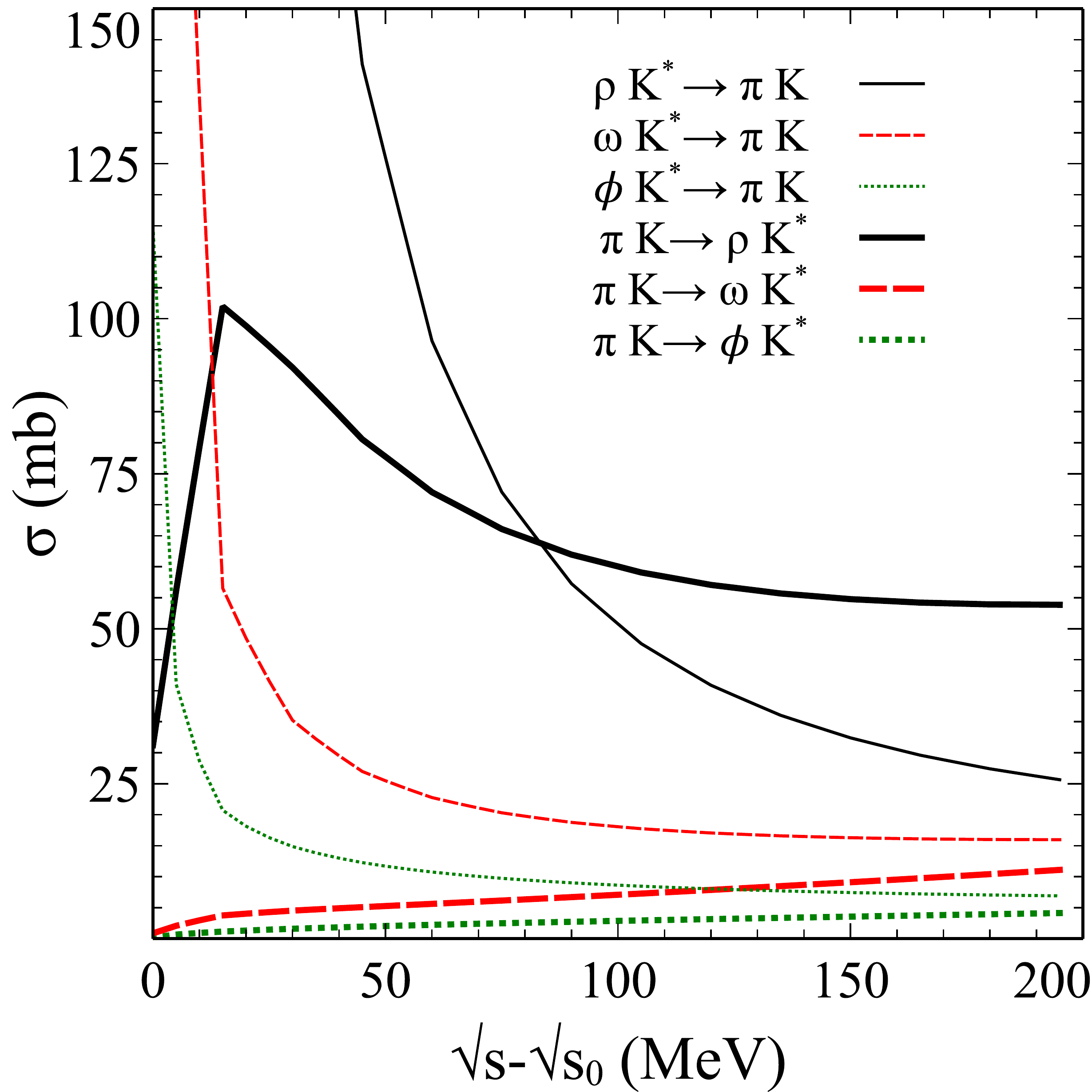} 
\caption{Cross sections for $\rho K,\omega K,\,\phi K\leftrightarrow \pi K^*$ (left panel) and 
$\pi K\leftrightarrow \rho K^*,\,\omega K^*,\, \phi K^*$ (right panel).}\label{PVtoPVandVVtoPP}
\end{figure}

We can obtain the cross section for the production of $K^*$ from the reactions $\rho K,\,\omega K,\,\phi K\to \pi K^*$ and $\pi K\to \rho K^*,\,\omega K^*,\,\phi K^*$ 
using the principle of detailed balance: if $\sigma_{ab\to cd}$ is the cross section for the process $a+b\to c+d$, calculated via 
Eq.~(\ref{cross}), we can determine the cross section for the inverse reaction, $c+d\to a+b$, as
\begin{align}
\sigma_{cd\to a b}=\frac{(2s_a+1)(2s_b+1)(2I_a+1)(2I_b+1)}{(2s_c+1)(2s_d+1)(2I_c+1)(2I_d+1)}
\frac{\lambda(s,m^2_a,m^2_b)}{\lambda(s,m^2_c,m^2_d)}\sigma_{ab\to cd}.\label{db}
\end{align}
In Fig.~\ref{PVtoPVandVVtoPP} we show the results obtained for the $K^*$ production cross sections using the principle of detailed balance. 
On the left panel, we show the cross sections found for $\rho K, \omega K, \phi K\to \pi K^*$ (thick lines) determined from the cross 
sections shown in Figs.~\ref{sigpiKstartorhoK},~\ref{sigpiKstartowphiK} (solid lines) and Eq.~(\ref{db}). The peak like structure found in the results for the reaction 
$\rho K\to \pi K^*$ corresponds to the manifestation of $K_1(1270)$. For the sake of comparison, we have also plotted the results found in 
Figs.~\ref{sigpiKstartorhoK},~\ref{sigpiKstartowphiK} for the cross sections of the $K^*$ absorption in the reactions $\pi K^*\to \rho K, \omega K, \phi K$. 
On the right panel, we show the results found for the cross sections of the reactions $\pi K\to \rho K^*,\omega K^*,\phi K^*$, which 
have been determined by using the results obtained in Figs.~\ref{sigrhoKstar},~\ref{sigwphiKstar},~\ref{sigTrLoopsrhoKstar},~\ref{sigTrLoopswphiKstar} (solid lines) and Eq.~(\ref{db}). 

As can be seen in Fig.~\ref{PVtoPVandVVtoPP} (left panel), the absorption cross sections of $K^*$ by $\pi$ are smaller than the corresponding 
ones for the production processes through collisions of $K$ with $\rho$, $\omega$ or $\phi$. The trend is the same in case of the absorption 
of $K^*$ by $\rho$ for excitation energies above $\sim$ 90 MeV (right panel), while the 
absorption cross sections of $K^*$ by $\omega$ or $\phi$ are larger 
than those related to its production from collisions of $\pi$ and $K$. However, for excitation energies bigger than $\sim90$ MeV the cross section for the $\pi K\to \rho K^*$ process dominates above all.

Very recently,  $K$ and $K^*$  formation in relativistic heavy-ion  
collisions has been investigated in the context of the Parton-Hadron-String  
dynamics (PHSD) transport approach~\cite{cabrera1,cabrera2}, which considers  
the in-medium effects in the $K$ and $\bar{K}^*$ states through the modification  
of their spectral properties during the propagation through the medium. The authors conclude that 
final state interactions (in the hadron gas) contribute to reduce the ratio $K^* / K$, corroborating 
the findings of \cite{cho}.

Our main results summarized in Fig.~\ref{PVtoPVandVVtoPP}.   
should be useful in the determination of the abundance ratio of $K^*$ and $K$ 
from heavy ion collisions with more accuracy. Based on these results, we may anticipate that, in contrast to 
the previous expectations \cite{cho,cabrera1,cabrera2}, the interactions in a pos-QGP hadronic medium 
may lead to an {\it enhancement} of the $K^*$ yield, not a suppression.  A detailed analysis based on the 
study of the rate equations is in progress and will be published soon. 

\section{Conclusions}
We have determined the cross sections related to the processes $\pi K^*\to \rho K,\,\omega K\, \phi K$ and 
$\rho K^*,\,\omega K^*\,\phi K^*\to \pi K$ considering the exchange of pseudoscalars, vectors and several 
resonances. The reactions $\pi K^*\to\rho K$ and $\rho K^*\to \pi K$, together with $K^*\to K\pi$, 
$K\pi\to K^*$, were found in Ref.~\cite{cho} to be the reactions contributing dominantly to the abundance 
ratio of $K^*$ and $K$ in heavy ion collisions.  However, several mechanisms which could contribute to the 
cross sections of $\pi K^*\to\rho K$ and $\rho K^*\to\pi K$ were missing in Ref.~\cite{cho}. With the purpose of 
obtaining information on such processes, we consider a more complete formalism, which takes into account more 
mechanisms  and calculate cross sections. We find that some of these 
new contributions turn out to be especially important, as the pseudoscalar exchange in the $t$-channel for the 
processes $\pi K^*\to \rho K$ and $\rho K^*\to \pi K$, exchange of resonances in the $s$-channel, like $K^*_2(1430)$, 
for $\rho K^*\to \pi K$, etc. We have also determined the cross sections for the inverse processes using the principle 
of detailed balance. The comparison between direct and inverse processes, shown in Fig.~\ref{PVtoPVandVVtoPP}, suggests that 
the production of $K^*$ in a hadron gas is more important that its absorption. 
Our results should be useful in obtaining a more accurate time evolution for the abundance ratio 
of $K^*$ and $K$ in heavy ion collisions.

\section{Acknowledgements} 
The authors would like to thank the Brazilian funding agencies FAPESP (under the grant number 2012/50984-4) and CNPq for the financial support (under the grant numbers 310759/2016-1 and 311524/2016-8).

\appendix
\section{Information related to the exchanged resonances in the $t$-, $u$- and $s$-channels}\label{apenA}
In Tables~\ref{K1} and \ref{RA} of this appendix we list, for completeness, the pole positions and couplings of the states found in Refs.~\cite{rocaS,gengR}. These couplings are in the isospin base and to determine their values in the charge basis we use the Clebsch-Gordan coefficients and the following convention to associate particles with states in the isospin base $|I,I_3\rangle$ (with $I$ being the total isospin and $I_3$ its third projection)
\begin{align}
|K^-\rangle=-\Big|\frac{1}{2},-\frac{1}{2}\Big\rangle,\quad |K^{*-}\rangle=-\Big|\frac{1}{2},-\frac{1}{2}\Big\rangle,\quad |\pi^+\rangle=-|1,1\rangle,\quad |\rho^+\rangle=-|1,1\rangle.
\end{align}
In this way, for example, from Table~\ref{RA}, we have that the coupling of the isospin 0, $G$-parity positive, state $f_1(1285)$ to $\frac{1}{\sqrt{2}}[\bar K^*K+K^*\bar K]$ (which corresponds to a positive $G$-parity combination) is $g=7230+i\,0$. This means that the state $f_1(1285)$ couples to the combination
\begin{align}
\frac{1}{\sqrt{2}}\left[|\bar K^*K,I=0,I_3=0\rangle+|K^*\bar K, I=0,I_3=0\rangle\right] & =\frac{1}{2}\left[|\bar K^{*0}K^0\rangle+|K^{*-}K^+\rangle \right. \nonumber \\
& \left. -|K^{*+}K^-\rangle-|K^{*0}\bar K^0\rangle\right],\nonumber
\end{align}
from which we get
\begin{align}
g_{f_1\to \bar K^{*0}K^0}=g_{f_1\to K^{*-}K^+}=-g_{f_1\to K^{*+}K^-}=-g_{f_1\to K^{*0}\bar K^0}=\frac{1}{2}g.
\end{align}

\begin{table}[h!]
\caption{Pole positions and couplings of the $K_1(1270)$ state to the different coupled channels whose dynamics generates the state~\cite{rocaS, gengR}. A two pole structure is found for $K_1(1270)$ in Refs.~\cite{rocaS, gengR} and the values shown in this table have been taken from Ref.~\cite{gengR}. The pole positions written in the table corresponds to $M-i\frac{\Gamma}{2}$, with $M$ and $\Gamma$ the mass and width characterizing the state.  The numerical values for the masses, widths and couplings of the states are expressed in MeV.}\label{K1}
\begin{tabular}{l|cc}

\text{Pole\,}&$1195-i\,123$&$1284-i\,73$\\
\hline
\text{Channel\,}&\multicolumn{2}{c}{\text{Coupling Constant}}\\
\hline
$\phi K$&$2096-i\,1208$&$1166-i\,774$\\
$\omega K$&$-2046+i\,821$&$-1051+i\,620$\\
$\rho K$&$-1671+i\,1599$&$4804+i\,395$\\
$K^*\eta$&$72+i\,197$&$3486-i\,536$\\
$K^*\pi$&$4747-i\,2874$&$769-i\,1171$\\
\end{tabular}
\end{table}

\begin{table}[h!]
\caption{Pole positions and couplings of the axial resonances $h_1(1170)$, $h_1(1380)$ (strangeness $S=0$, isospin $I=0$, $G$-parity $G=-$), $f_1(1285)$ ($S=0$, $I=0$, $G=+$), $a_1(1260)$ ($S=0$, $I=1$, $G=-$) and $b_1(1235)$ ($S=0$, $I=1$, $G=+$) to the different coupled channels whose dynamics generates them~\cite{rocaS}. The values shown in this table have been taken from Ref.~\cite{rocaS}. The pole positions written in the table corresponds to $M-i\frac{\Gamma}{2}$, with $M$ and $\Gamma$ the mass and width characterizing the state. The numerical values for the masses, widths and couplings of the states are expressed in MeV.}\label{RA}
\begin{tabular}{l|ccc}
\text{State ($G=-$)}&$h_1(1170)$&$h_1(1380)$&$a_1(1260)$\\
\text{Pole\,}&$919-i\,17$&$1245-i\,7$&$1011-i84$\\
\hline
\text{Channel ($G=-$)\,}&\multicolumn{3}{c}{\text{Coupling Constant}}\\
\hline
$\frac{1}{\sqrt{2}}[\bar K^*K-K^*\bar K]$&$781-i\,498$&$6147+i\,183$&$1872-i\,1486$\\
$\phi \eta$&$46-i\,13$&$-3311+i\,47$&$0$\\
$\omega \eta$&$23-i\,28$&$3020-i\,22$&$0$\\
$\rho\pi$&$-3453+i\,1681$&$648-i\,959$&$-3795+i\,2330$\\
\end{tabular}
\\
\vspace{1cm}
\begin{tabular}{l|cc}
\text{State ($G=+$)}&$f_1(1285)$&$b_1(1235)$\\
\text{Pole\,}&$1288-i\,0$&$1245-i\,28$\\
\hline
\text{Channel ($G=+$)\,}&\multicolumn{2}{c}{\text{Coupling Constant}}\\
\hline
$\frac{1}{\sqrt{2}}[\bar K^*K+K^*\bar K]$&$7230+i\,0$&$6172-i\,75$\\
$\phi \pi$&$0$&$2087-i\,385$\\
$\omega \pi$&$0$&$-1869+i\,300$\\
$\rho\eta$&$0$&$-3041+i\,498$\\
\end{tabular}

\end{table}

In Tables~\ref{CijT}-\ref{CijSrho} we give the coefficients entering in the $t$-, $u$- and $s$- channel amplitudes of Figs.~\ref{piKstar} and~\ref{rhoKstar}.

\begin{table}
\caption{Coefficients $\mathbb{T}^{\,ij}_k$ present in Eq.~(\ref{Tpi}) for the reactions $r=$ $\pi K^*\to \rho K,\, \omega K,\,\phi K$ for total charge $-1$, $0$, $1$ and $2$. The index $i$ represents the initial state $\pi K^*$ for a particular charge configuration and the index $j$ corresponds to the final state for the same total charge. The index $k$ corresponds to the exchanged pseudoscalar, which we indicate in brackets next to the coefficient. The absence of the coefficient for some $k$ means that the coefficient is 0 for that exchanged particle. If no exchanged particle is indicated next to the coefficient, the coefficient is 0 independently of the exchanged particle.}\label{CijT}

\begin{tabular}{c|c}
&$\rho^- K^0$\\
\hline
$\pi^-K^{*0}$&$4\,(\pi^0)$
\end{tabular}
\hspace{0.5cm}
\begin{tabular}{c|cccc}
&$\rho^- K^+$&$\rho^0 K^0$&$\omega K^0$&$\phi K^0$\\
\hline
$\pi^-K^{*+}$&$-4\,(\pi^0)$&$4\sqrt{2}\,(\pi^-)$&$0$&$0$\\
$\pi^0 K^{*0}$&$4\sqrt{2}\,(\pi^+)$&$0$&$0$&$0$
\end{tabular}
\\
\vspace{0.5cm}
\begin{tabular}{c|c}
&$\rho^+ K^+$\\
\hline
$\pi^+K^{*+}$&$4\,(\pi^0)$
\end{tabular}
\hspace{0.5cm}
\begin{tabular}{c|cccc}
&$\rho^0 K^+$&$\rho^+ K^0$&$\omega K^+$&$\phi K^+$\\
\hline
$\pi^0K^{*+}$&$0$&$-4\sqrt{2}\,(\pi^-)$&$0$&$0$\\
$\pi^+ K^{*0}$&$-4\sqrt{2}\,(\pi^+)$&$-4\,(\pi^0)$&$0$&$0$
\end{tabular}
\end{table}

\begin{table}
\caption{Coefficients $\mathbb{\overline{T}}^{\,ij}_k$ present in Eq.~(\ref{Tpi}) for the reactions $r=$ $\pi K^*\to \rho K,\, \omega K,\,\phi K$ for total charge $-1$, $0$, $1$ and $2$. The index $i$ represents the initial state $\pi K^*$ for a particular charge configuration and the index $j$ corresponds to the final state for the same total charge. The index $k$ corresponds to the exchanged vector, which we indicate in brackets next to the coefficient. See the caption of Table~\ref{CijT} for the notation used here.}\label{CijbarT}

\begin{tabular}{c|c}
&$\rho^- K^0$\\
\hline
$\pi^-K^{*0}$&$-\frac{1}{2}\,(\omega)$
\end{tabular}
\hspace{0.5cm}
\begin{tabular}{c|cccc}
&$\rho^- K^+$&$\rho^0 K^0$&$\omega K^0$&$\phi K^0$\\
\hline
$\pi^-K^{*+}$&$-\frac{1}{2}\,(\omega)$&0&$-\frac{1}{\sqrt{2}}\,(\rho^-)$&$0$\\
$\pi^0 K^{*0}$&$0$&$-\frac{1}{2}\,(\omega)$&$\frac{1}{2}\,(\rho^0)$&$0$
\end{tabular}
\\
\vspace{0.5cm}
\begin{tabular}{c|c}
&$\rho^+ K^+$\\
\hline
$\pi^+K^{*+}$&$-\frac{1}{2}\,(\omega)$
\end{tabular}
\hspace{0.5cm}
\begin{tabular}{c|cccc}
&$\rho^0 K^+$&$\rho^+ K^0$&$\omega K^+$&$\phi K^+$\\
\hline
$\pi^0K^{*+}$&$-\frac{1}{2}\,(\omega)$&0&$-\frac{1}{2}\,(\rho^0)$&$0$\\
$\pi^+ K^{*0}$&$0$&$-\frac{1}{2}\,(\omega)$&$-\frac{1}{\sqrt{2}}\,(\rho^+)$&$0$
\end{tabular}
\end{table}

\begin{table}
\caption{Coefficients $\mathbb{U}^{\,ij}_k$ present in Eq.~(\ref{Upi}) for the reactions $r=$ $\pi K^*\to \rho K,\, \omega K,\,\phi K$ for total charge $-1$, $0$, $1$ and $2$. See the caption of Table~\ref{CijbarT} for the notation used here.}\label{CijU}
\begin{tabular}{c|c}
&$\rho^- K^0$\\
\hline
$\pi^-K^{*0}$&$1\,(K^{*-})$
\end{tabular}
\hspace{0.5cm}
\begin{tabular}{c|cccc}
&$\rho^- K^+$&$\rho^0 K^0$&$\omega K^0$&$\phi K^0$\\
\hline
$\pi^-K^{*+}$&$0$&$\frac{1}{\sqrt{2}}\,(K^{*-})$&$\frac{1}{\sqrt{2}}\,(K^{*-})$&$-1\,(K^{*-})$\\
$\pi^0 K^{*0}$&$\frac{1}{\sqrt{2}}\,(K^{*-})$&$\frac{1}{2}\,(\bar K^{*0})$&$-\frac{1}{2}\,(\bar K^{*0})$&$\frac{1}{\sqrt{2}}\,(\bar K^{*0})$
\end{tabular}
\\
\vspace{0.5cm}
\begin{tabular}{c|c}
&$\rho^+ K^+$\\
\hline
$\pi^+K^{*+}$&$1\,(\bar K^{*0})$
\end{tabular}
\hspace{0.5cm}
\begin{tabular}{c|cccc}
&$\rho^0 K^+$&$\rho^+ K^0$&$\omega K^+$&$\phi K^+$\\
\hline
$\pi^0K^{*+}$&$\frac{1}{2}\,(K^{*-})$&$-\frac{1}{\sqrt{2}}\,(\bar K^{*0})$&$\frac{1}{2}\,(K^{*-})$&$-\frac{1}{\sqrt{2}}\,(K^{*-})$\\
$\pi^+ K^{*0}$&$-\frac{1}{\sqrt{2}}\,(\bar K^{*0})$&$0$&$\frac{1}{\sqrt{2}}\,(\bar K^{*0})$&$-1\,(\bar K^{*0})$
\end{tabular}

\end{table}

\begin{table}
\caption{Coefficients $\mathbb{S}^{\,ij}_K$ present in Eq.~(\ref{Spi}) for the reactions $r=$ $\pi K^*\to \rho K,\, \omega K,\,\phi K$ for total charge $-1$, $0$, $1$ and $2$. In this case, a $K^0$ is exchanged for those processes whose total charge is $0$ and a $K^+$ for total charge $+1$. For total charge $-1$ and $2$, no particle can be exchanged in the $s$-channel.}\label{CijSK}
\begin{tabular}{c|c}
&$\rho^- K^0$\\
\hline
$\pi^-K^{*0}$&$0$
\end{tabular}
\hspace{0.5cm}
\begin{tabular}{c|cccc}
&$\rho^- K^+$&$\rho^0 K^0$&$\omega K^0$&$\phi K^0$\\
\hline
$\pi^-K^{*+}$&$-4$&$2\sqrt{2}$&$-2\sqrt{2}$&$4$\\
$\pi^0 K^{*0}$&$2\sqrt{2}$&$-2$&$2$&$-2\sqrt{2}$
\end{tabular}
\\
\begin{tabular}{c|c}
&$\rho^+ K^+$\\
\hline
$\pi^+K^{*+}$&$0$
\end{tabular}
\hspace{0.5cm}
\begin{tabular}{c|cccc}
&$\rho^0 K^+$&$\rho^+ K^0$&$\omega K^+$&$\phi K^+$\\
\hline
$\pi^0K^{*+}$&$-2$&$-2\sqrt{2}$&$-2$&$2\sqrt{2}$\\
$\pi^+ K^{*0}$&$-2\sqrt{2}$&$-4$&$-2\sqrt{2}$&$4$
\end{tabular}

\end{table}

\begin{table}
\caption{Coefficients $\mathbb{S}^{\,ij}_{K^*}$ present in Eq.~(\ref{Spi}) for the reactions $r=$ $\pi K^*\to \rho K,\, \omega K,\,\phi K$ for total charge $-1$, $0$, $1$ and $2$. In this case, a $K^{*0}$ is exchanged for those processes whose total charge is $0$ and a $K^{*+}$ for total charge $+1$. In case of total charge $-1$ and $2$, no particle can be exchanged in the $s$-channel.}\label{CijSKstar}
\begin{tabular}{c|c}
&$\rho^- K^0$\\
\hline
$\pi^-K^{*0}$&$0$
\end{tabular}
\hspace{0.5cm}
\begin{tabular}{c|cccc}
&$\rho^- K^+$&$\rho^0 K^0$&$\omega K^0$&$\phi K^0$\\
\hline
$\pi^-K^{*+}$&$-\frac{1}{2}$&$\frac{1}{2\sqrt{2}}$&$-\frac{1}{2\sqrt{2}}$&$-\frac{1}{2}$\\
$\pi^0 K^{*0}$&$\frac{1}{2\sqrt{2}}$&$-\frac{1}{4}$&$\frac{1}{4}$&$\frac{1}{2\sqrt{2}}$
\end{tabular}
\\
\vspace{0.5cm}
\begin{tabular}{c|c}
&$\rho^+ K^+$\\
\hline
$\pi^+K^{*+}$&$0$
\end{tabular}
\hspace{0.5cm}
\begin{tabular}{c|cccc}
&$\rho^0 K^+$&$\rho^+ K^0$&$\omega K^+$&$\phi K^+$\\
\hline
$\pi^0K^{*+}$&$-\frac{1}{4}$&$-\frac{1}{2\sqrt{2}}$&$-\frac{1}{4}$&$-\frac{1}{2\sqrt{2}}$\\
$\pi^+ K^{*0}$&$-\frac{1}{2\sqrt{2}}$&$-\frac{1}{2}$&$-\frac{1}{2\sqrt{2}}$&$-\frac{1}{2}$
\end{tabular}

\end{table}

\begin{table}
\caption{Coefficients $\mathcal{T}^{\,ij}$ present in Eq.~(\ref{Trho}) for the reactions $r=$ $\rho K^*,\, \omega K^*,\,\phi K^*\to\pi K$ for total charge $-1$, $0$, $1$ and $2$. In this case, a vector meson is exchanged and we write the exchanged particle next to the coefficient. If the coefficient is 0 the process can not proceed via vector meson exchange.}\label{CijTrho}
\begin{tabular}{c|c}
&$\pi^- K^0$\\
\hline
$\rho^-K^{*0}$&$4\,(\pi^0)$
\end{tabular}
\hspace{0.5cm}
\begin{tabular}{c|cc}
&$\pi^- K^+$&$\pi^0 K^0$\\
\hline
$\rho^-K^{*+}$&$-4\,(\pi^0)$&$4\sqrt{2}\,(\pi^-)$\\
$\rho^0 K^{*0}$&$4\sqrt{2}\,(\pi^+)$&$0$\\
$\omega K^{*0}$&$0$&$0$\\
$\phi K^{*0}$&$0$&$0$
\end{tabular}
\\
\vspace{0.5cm}
\begin{tabular}{c|c}
&$\pi^+ K^+$\\
\hline
$\rho^+K^{*+}$&$4\,(\pi^0)$
\end{tabular}
\hspace{0.5cm}
\begin{tabular}{c|cc}
&$\pi^0 K^+$&$\pi^+ K^0$\\
\hline
$\rho^0K^{*+}$&$0$&$-4\sqrt{2}\,(\pi^-)$\\
$\rho^+ K^{*0}$&$-4\sqrt{2}\,(\pi^+)$&$-4\,(\pi^0)$\\
$\omega K^{*+}$&$0$&$0$\\
$\phi K^{*+}$&$0$&$0$\\
\end{tabular}

\end{table}

\begin{table}
\caption{Coefficients $\mathcal{\overline{T}}^{\,ij}$ present in Eq.~(\ref{Trho}) for the reactions $r=$ $\rho K^*,\, \omega K^*,\,\phi K^*\to\pi K$ for total charge $-1$, $0$, $1$ and $2$. See the caption of Table~\ref{CijTrho} for the notation used here.}\label{CijbarTrho}
\begin{tabular}{c|c}
&$\pi^- K^0$\\
\hline
$\rho^-K^{*0}$&$-\frac{1}{2}\,(\omega)$
\end{tabular}
\hspace{0.5cm}
\begin{tabular}{c|cc}
&$\pi^- K^+$&$\pi^0 K^0$\\
\hline
$\rho^-K^{*+}$&$-\frac{1}{2}\,(\omega)$&$0$\\
$\rho^0 K^{*0}$&$0$&$-\frac{1}{2}\,(\omega)$\\
$\omega K^{*0}$&$-\frac{1}{\sqrt{2}}\,(\rho^+)$&$\frac{1}{2}\,(\rho^0)$\\
$\phi K^{*0}$&$0$&$0$
\end{tabular}
\\
\vspace{0.5cm}
\begin{tabular}{c|c}
&$\pi^+ K^+$\\
\hline
$\rho^+K^{*+}$&$-\frac{1}{2}\,(\omega)$
\end{tabular}
\hspace{0.5cm}
\begin{tabular}{c|cc}
&$\pi^0 K^+$&$\pi^+ K^0$\\
\hline
$\rho^0K^{*+}$&$-\frac{1}{2}\,(\omega)$&$0$\\
$\rho^+ K^{*0}$&$0$&$-\frac{1}{2}\,(\omega)$\\
$\omega K^{*+}$&$-\frac{1}{2}\,(\rho^0)$&$-\frac{1}{\sqrt{2}}\,(\rho^-)$\\
$\phi K^{*+}$&$0$&$0$\\
\end{tabular}

\end{table}

\begin{table}
\caption{Coefficients $\mathcal{U}^{\,ij}_{\bar{K}}$ present in Eq.~(\ref{Urho}) for the reactions $r=$ $\rho K^*,\, \omega K^*,\,\phi K^*\to\pi K$ for total charge $-1$, $0$, $1$ and $2$. In this case, a $\bar K$ meson is exchanged in all diagrams.}\label{CijUrho}
\begin{tabular}{c|c}
&$\pi^- K^0$\\
\hline
$\rho^-K^{*0}$&$4$
\end{tabular}
\hspace{0.5cm}
\begin{tabular}{c|cc}
&$\pi^- K^+$&$\pi^0 K^0$\\
\hline
$\rho^-K^{*+}$&$0$&$2\sqrt{2}$\\
$\rho^0 K^{*0}$&$2\sqrt{2}$&$2$\\
$\omega K^{*0}$&$2\sqrt{2}$&$-2$\\
$\phi K^{*0}$&$-4$&$2\sqrt{2}$
\end{tabular}
\\
\vspace{0.5cm}
\begin{tabular}{c|c}
&$\pi^+ K^+$\\
\hline
$\rho^+K^{*+}$&$4$
\end{tabular}
\hspace{0.5cm}
\begin{tabular}{c|cc}
&$\pi^0 K^+$&$\pi^+ K^0$\\
\hline
$\rho^0K^{*+}$&$2$&$-2\sqrt{2}$\\
$\rho^+ K^{*0}$&$-2\sqrt{2}$&$0$\\
$\omega K^{*+}$&$2$&$2\sqrt{2}$\\
$\phi K^{*+}$&$-2\sqrt{2}$&$-4$\\
\end{tabular}

\end{table}

\begin{table}
\caption{Coefficients $\mathcal{U}^{\,ij}_{\bar K^*}$ present in Eq.~(\ref{Urho}) for the reactions $r=$ $\rho K^*,\, \omega K^*,\,\phi K^*\to\pi K$ for total charge $-1$, $0$, $1$ and $2$. In this case, a $\bar K^*$ meson is exchanged in all diagrams.}\label{CijUrho2}
\begin{tabular}{c|c}
&$\pi^- K^0$\\
\hline
$\rho^-K^{*0}$&$-\frac{1}{2}$
\end{tabular}
\hspace{0.5cm}
\begin{tabular}{c|cc}
&$\pi^- K^+$&$\pi^0 K^0$\\
\hline
$\rho^-K^{*+}$&$0$&$-\frac{1}{2\sqrt{2}}$\\
$\rho^0 K^{*0}$&$-\frac{1}{2\sqrt{2}}$&$-\frac{1}{4}$\\
$\omega K^{*0}$&$-\frac{1}{2\sqrt{2}}$&$\frac{1}{4}$\\
$\phi K^{*0}$&$-\frac{1}{2}$&$\frac{1}{2\sqrt{2}}$
\end{tabular}
\\
\vspace{0.5cm}
\begin{tabular}{c|c}
&$\pi^+ K^+$\\
\hline
$\rho^+K^{*+}$&$-\frac{1}{2}$
\end{tabular}
\hspace{0.5cm}
\begin{tabular}{c|cc}
&$\pi^0 K^+$&$\pi^+ K^0$\\
\hline
$\rho^0K^{*+}$&$-\frac{1}{4}$&$\frac{1}{2\sqrt{2}}$\\
$\rho^+ K^{*0}$&$\frac{1}{2\sqrt{2}}$&$0$\\
$\omega K^{*+}$&$-\frac{1}{4}$&$-\frac{1}{2\sqrt{2}}$\\
$\phi K^{*+}$&$-\frac{1}{2\sqrt{2}}$&$-\frac{1}{2}$\\
\end{tabular}

\end{table}

\begin{table}
\caption{Coefficients $\mathcal{S}^{\,ij}$ present in Eq.~(\ref{Srho}) for the reactions $r=$ $\rho K^*,\, \omega K^*,\,\phi K^*\to\pi K$ for total charge $-1$, $0$, $1$ and $2$. In this case, a $\bar K^*$ meson is exchanged in all diagrams.}\label{CijSrho}
\begin{tabular}{c|c}
&$\pi^- K^0$\\
\hline
$\rho^-K^{*0}$&$0$
\end{tabular}
\hspace{0.5cm}
\begin{tabular}{c|cc}
&$\pi^- K^+$&$\pi^0 K^0$\\
\hline
$\rho^-K^{*+}$&$1$&$-\frac{1}{\sqrt{2}}$\\
$\rho^0 K^{*0}$&$-\frac{1}{\sqrt{2}}$&$\frac{1}{2}$\\
$\omega K^{*0}$&$\frac{1}{\sqrt{2}}$&$-\frac{1}{2}$\\
$\phi K^{*0}$&$-1$&$\frac{1}{\sqrt{2}}$
\end{tabular}
\\
\vspace{0.5cm}
\begin{tabular}{c|c}
&$\pi^+ K^+$\\
\hline
$\rho^+K^{*+}$&$0$
\end{tabular}
\hspace{0.5cm}
\begin{tabular}{c|cc}
&$\pi^0 K^+$&$\pi^+ K^0$\\
\hline
$\rho^0K^{*+}$&$\frac{1}{2}$&$\frac{1}{\sqrt{2}}$\\
$\rho^+ K^{*0}$&$\frac{1}{\sqrt{2}}$&$1$\\
$\omega K^{*+}$&$\frac{1}{2}$&$\frac{1}{\sqrt{2}}$\\
$\phi K^{*+}$&$-\frac{1}{\sqrt{2}}$&$-1$\\
\end{tabular}

\end{table}

\section{Evaluation of the $s$-channel exchange of resonances in the reactions $\rho K^*, \omega K^*, \phi K^*\to \pi K$}\label{apenB}
In this appendix, we determine the amplitude related to the process depicted in Fig.~\ref{res_sch_exch} in which the $K^*_S$ states (where the subscript $S$ indicates spin) found in Ref.~\cite{gengo} are exchanged in the $s$-channel through triangular loops (see Fig.~\ref{square}). We have summarized the properties found in Ref.~\cite{gengo} for these $K^*_S$ in Table~\ref{RV}. 

\begin{table}[h!]
\caption{Pole positions and couplings of the vector resonances $K^*_0 (1643)$ ($I=1/2$, spin 0) and $K^*_2(1430)$ ($I=1/2$, spin 2) found in Ref.~\cite{gengo}, with the former being a prediction of the model. The pole positions written in the table corresponds to $M-i\frac{\Gamma}{2}$, with $M$ and $\Gamma$ the mass and width characterizing the state. The numerical values for the masses, widths and couplings of the states are expressed in MeV.}\label{RV}
\begin{tabular}{l|cc}
\text{State}&$K^*_0(1643)$&$K^*_2(1430)$\\
\text{Pole\,}&$1643-i\,24$&$1431-i\,28$\\
\hline
\text{Channel\,}&\multicolumn{2}{c}{\text{Coupling Constant}}\\
\hline
$\rho K^*$&$8102-i\,959$&$10901-i\,71$\\
$\omega K^*$&$1370-i\,146$&$2267-i\,13$\\
$\phi K^*$&$-1518+i\,209$&$-2898+i\,17$\\
\end{tabular}
\end{table}

We have the following expression for the amplitude of the process depicted in Fig.~\ref{res_sch_exch}
\begin{align}
\pmb{S}^{ij}_{K^*_S}&=\sum_{k_1,k_2,k_3}g^{(i)}_{K^*_S}g^{(k_1k_2)}_{K^*_S}\frac{1}{s-M^2_{K^*_S}+i\,\Gamma_{K^*_S}M_{K^*_S}}g^2_{PPV}\mathcal{S}^{(k_1k_2k_3)} P^{\mu\nu}_S p^{\prime}_\mu I^{(k_1k_2k_3)}_\nu,\label{SKstar}
\end{align}
where the coefficients $\mathcal{S}^{(k_1k_2k_3)}$ are given in Table~\ref{CijS12}. The symbols $M_{K^*_S}$ and $\Gamma_{K^*_S}$ in Eq.~(\ref{SKstar}) are the mass and width of the poles related to the exchanged $K^*_S$ state, while $g^{(i)}_{K^*_S}$ and $g^{(k_1k_2)}_{K^*_S}$ are, respectively, the coupling constants of those poles to the initial state and to the vector mesons present in the triangular loops shown in Fig.~\ref{square}. The numerical values for these quantities can be found in Table~\ref{RV}.

\begin{table}
\caption{Coefficients $\mathcal{S}^{(k_1k_2k_3)}$ present in Eq.~(\ref{SKstar}) for the reactions $r=$ $\rho K^*,\, \omega K^*,\,\phi K^*\to\pi K$ for total charge $-1$, $0$, $1$ and $2$. We indicate those particles (related to the indices $k_1$, $k_2$ and $k_3$) which, when involved in the triangular loop, give a nonzero coefficient.}\label{CijS12}
\begin{tabular}{c|c}
$(k_1k_2k_3)$&$\mathcal{S}^{(k_1k_2k_3)}$ \\
\hline
$\rho^-K^{*+}\pi^0$&$2$\\
$\rho^0 K^{*0}\pi^+$&$-2\sqrt{2}$\\
$\rho^-K^{*+}\pi^-$&$-2\sqrt{2}$\\
$\rho^+K^{*0}\pi^+$&$2\sqrt{2}$\\
$\rho^0 K^{*+}\pi^-$&$2\sqrt{2}$\\
$\rho^+K^{*0}\pi^0$&$2$
\end{tabular}
\end{table}

To get Eq.~(\ref{SKstar}), we have used the following amplitude for the coupling of the $K^*_S$ states to the vector mesons 
\begin{align}
t_{K^*_S}=g^{(i)}_{K^*_S}g^{(k_1k_2)}_{K^*_S}\frac{1}{s-M^2_{K^*_S}+i\Gamma_{K^*_S}M_{K^*_S}}P_S,
\end{align}
where $P_S$ is a spin projector, which is given for the case of spin $S=0,2$ by~\cite{raquel}
\begin{align}
P_0&=\frac{1}{3}\epsilon_\mu(k)\epsilon^\mu(p)\epsilon_\nu(q)\epsilon^\nu(p+k-q),\nonumber\\
P_2&=\frac{1}{2}[\epsilon_\mu(p)\epsilon_\nu(k)\epsilon^\mu(q)\epsilon^\nu(p+k-q)+\epsilon_\mu(p)\epsilon_\nu(k)\epsilon^\nu(q)\epsilon^\mu(p+k-q)]\nonumber\\
&\quad-\frac{1}{3}\epsilon_\alpha(p)\epsilon^\alpha(k)\epsilon_\beta(q)\epsilon^\beta(p+k-q).\label{pro}
\end{align}
In Eq.~(\ref{pro}), $q$ and $p+k-q$ represent, respectively, the four momenta of the vector meson without strangeness and the $K^*$  meson present in the triangular loop of Fig.~\ref{square} and which are coupled to $K^*_S$.

Since $K^*_S$ can be considered as molecular state of $\rho K^*$ and coupled channels~\cite{gengo} with its hadron components being in $s$-wave, the vector mesons present in the triangular loops and which couple to $K^*_S$, although being off-shell, should not be very far from being on-shell (i.e., their respective modulus of the three-momenta are negligible as compared to their energies). Within such an interpretation of $K^*_S$, the temporal part of the polarization vectors ($\sim$ modulus of momentum divided by mass) of the mesons at the resonance-meson-meson vertex  should be negligible as compared to the spatial components. This means that for the external as well as the internal vector mesons coupled to $K^*_S$ we can use the approximation~\cite{gengo,raquel}
\begin{align}
\sum_\text{polarizations}\epsilon^\mu\epsilon^\nu\sim \sum_\text{polarizations}\epsilon^i\epsilon^j=\delta^{ij},\label{aproxeo}
\end{align}
with $i$ and $j$ being spatial indices. However, it would be more appropriate to maintain the covariant formalism instead of working with mixed indices (some spatial and other temporal-spatial). This can be achieved by writting
\begin{align}
\sum_\text{polarizations}\epsilon^\mu\epsilon^\nu\sim -g^{\mu\nu},\label{aprox1}
\end{align}
for the vector mesons coupled to $K^*_S$ present in the triangular loop of Fig.~\ref{square}. This approximation implies the inclusion, in the result, of a very small contribution arising from the temporal part of the polarization vector of these vector mesons.  We have made use of this approximation to get Eq.~(\ref{SKstar}). When summing over the polarizations of the external vector mesons coupled to $K^*_S$ we use
\begin{align}
\sum_\text{polarizations}\epsilon^\mu(k)\epsilon^\nu(k)&=-g^{\mu\nu}+\frac{k^\mu k^\nu}{m^2_{K^*}},\nonumber\\
\sum_\text{polarizations}\epsilon^\mu(p)\epsilon^\nu(p)&=-g^{\mu\nu}+\frac{p^\mu p^\nu}{m^2_V},\label{aprox2}
\end{align}
which will produce negligible values for the temporal and temporal-spatial components. This is so because, as mentioned above, the external vectors, when interacting in $s$-wave and for energies not far away from the threshold (as in our case), generate the $K^*_S$ (following the interpretation of Ref.~\cite{gengo}). Thus, the modulus of their momenta is much smaller than their energies, so
\begin{align}
\sum_\text{polarizations}\epsilon^0(k)\epsilon^0(k)=-g^{00}+\frac{k^0 k^0}{m^2_{K^*}}&=-1+\frac{k^0 k^0}{m^2_{K^*}}\sim-1+1=0,\nonumber\\
\sum_\text{polarizations}\epsilon^i(k)\epsilon^0(k)=-g^{i0}+\frac{k^i k^0}{m^2_{K^*}}&=\frac{k^i k^0}{m^2_{K^*}}\sim 0,\nonumber\\
 \sum_\text{polarizations}\epsilon^i(k)\epsilon^j(k)-g^{ij}+\frac{k^i k^j}{m^2_{K^*}}&=1+\frac{k^i k^j}{m^2_{K^*}}\sim 1,
\end{align}
and same is the case for $\epsilon(p)$. Then, the use of Eqs.~(\ref{aprox1}) and (\ref{aprox2}) is in line with the approximation in Eq.~(\ref{aproxeo}).

The summation over the polarizations of the vector mesons in the triangular loop coupled to $K^*_S$ gives rise to the $P^{\mu\nu}_S$ present in Eq.~(\ref{SKstar}), which is a spin projector for the external vector mesons coupled to $K^*_S$. Within the approximation of Eqs.~(\ref{aprox1}) and ~(\ref{aprox2}), we have for spin $S=0,2$ 
\begin{align}
P^{\mu\nu}_0&=\frac{1}{3}\epsilon(p)\cdot \epsilon(k) g^{\mu\nu},\nonumber\\
P^{\mu\nu}_2&=\frac{1}{2}[\epsilon_\mu(p)\epsilon_\nu(k)+\epsilon_\nu(p)\epsilon_\mu(k)]-\frac{1}{3}\epsilon(p)\cdot\epsilon(k)g^{\mu\nu}.
\end{align}
These expressions can be compared with the spin projectors found in Ref.~\cite{raquel} for the case of spatial indices and neglecting the temporal part of the polarization vector, 
\begin{align}
P^{ij}_0&=\frac{1}{3}\vec{\epsilon}(p)\cdot \vec{\epsilon}(k)\delta^{ij} ,\nonumber\\
P^{ij}_2&=\frac{1}{2}[\epsilon_i(p)\epsilon_j(k)+\epsilon_j(p)\epsilon_i(k)]-\frac{1}{3}\vec{\epsilon}(p)\cdot\vec{\epsilon}(k)\delta^{ij}.
\end{align}
In this case, Eq.~(\ref{aproxeo}) is used to sum over the polarizations.

\begin{figure}[h!]
\centering
\includegraphics[width=0.6\textwidth]{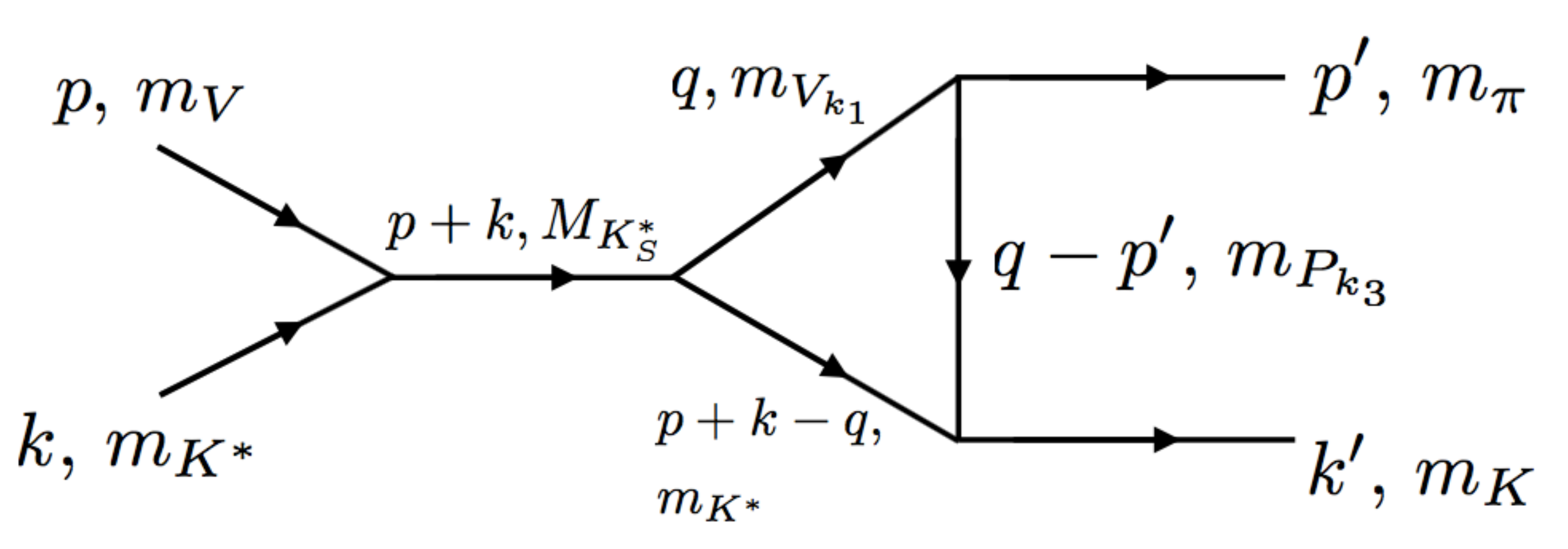}
\caption{Momentum and mass assignations for the particles involved in the triangular loop diagram of Fig.~\ref{square}. The mass $m_V$ is associated with the external $\rho$, $\omega$ or $\phi$ mesons; $m_{V_{k_1}}$ and $m_{V_{k_2}}=m_{K^*}$ are the masses of the vector mesons which can couple to $K^*_S$ and they are listed in Table~\ref{RV}. The mass $m_{P_{k_3}}$ is related to the pseudoescalars ($\pi$, $\eta$, $\eta^\prime$) which can be exchanged.}. \label{Tloop}
\end{figure}

In Eq.~(\ref{SKstar}), $I^{(k_1k_2k_3)}_\nu$ corresponds to the following integral
\begin{align}
I^{(k_1k_2k_3)}_\nu&=\int \frac{d^4q}{(2\pi)^4}\frac{1}{q^2-m^2_{V_{k_1}}+i\epsilon}\frac{(k^\prime-p^\prime+q)_\nu}{(p+k-q)^2-m^2_{V_{k_2}}+i\epsilon}\frac{1}{(q-p^\prime)^2-m^2_{P_{k_3}}+i\epsilon},\label{Inu}
\end{align}
with $m_{V_{k_1}}$, $m_{V_{k_2}}=m_{K^*}$ being the masses of the two vector mesons which couple to $K^*_S$ in the triangular loop of Fig.~\ref{square} and $m_{P_{k_3}}$ is the mass of the exchanged pseudoscalar. Using Lorentz covariance, the integral of Eqs.~(\ref{Inu}) can be written as 
\begin{align}
I^{(k_1k_2k_3)}_\nu&=a^{(k_1k_2k_3)} k_\nu+b^{(k_1k_2k_3)}p_\nu+c^{(k_1k_2k_3)}k^\prime_\nu+d^{(k_1k_2k_3)}p^\prime_\nu,\label{IL}
\end{align}
and we need to determine the coefficients $a^{(k_1k_2k_3)}, b^{(k_1k_2k_3)}, \cdots$ appearing in this expression. The momentum and mass assignations for the particles involved in the triangular loop diagrams is shown in Fig.~\ref{Tloop}. 

The determination of the four coefficients of Eq.~(\ref{IL}) can be done by making use of the Feynman parametrization and writing
\begin{align}
\frac{1}{\alpha\beta\gamma}=2\int_0^1 dx\int_0^x dy\frac{1}{[\alpha+(\beta-\alpha)x+(\gamma-\beta)y]^3},\label{feynp}
\end{align}
where 
\begin{align}
\alpha\equiv q^2-m^2_{V_{k_1}}, \quad \beta\equiv (p+k-q)^2-m^2_{K^*},\quad\gamma=(q-p^\prime)^2-m^2_{P_{k_3}}.
\end{align}
In this way,
\begin{align}
[\alpha+(\beta-\alpha)x+(\gamma-\beta)y]=q^{\prime\,2}+r^{(k_1k_2k_3)},\label{rel}
\end{align}
where we have defined 
\begin{align}
q^\prime&\equiv q-(p+k)(x-y)+p^\prime y,\label{qprime}
\end{align}
and
\begin{align}
r^{(k_1k_2k_3)}&= -(x-y)\left[(m^2_V+m^2_{K^*}+2p\cdot k)(x-y-1)+2p^\prime\cdot(p+k)y+m^2_{V_{k_2}}\right]\nonumber\\
&\quad+[(1-y)m^2_\pi-m^2_{P_{k_3}}]y+m^2_{V_{k_1}}(x-1)\label{r1}
\end{align}

Using Eqs.~(\ref{feynp}),~(\ref{rel}),~(\ref{qprime}),~(\ref{r1}), and the relation
\begin{align}
\int \frac{d^4q^\prime}{(2\pi)^4}\frac{1}{(q^{\prime\,2}+r+i\epsilon)^3}=\frac{i}{2^5\pi^2(r+i\epsilon)},
\end{align}
we can identify the coefficients in Eqs.~(\ref{IL}),
\begin{align}
a^{(k_1k_2k_3)}&=b^{(k_1k_2k_3)}=\frac{1}{2^4\pi^2}\int_0^1 dx\int_0^x dy\frac{(x-y)}{r^{(k_1k_2k_3)}+i\epsilon},\nonumber\\
c^{(k_1k_2k_3)}&=\frac{1}{2^4\pi^2}\int_0^1 dx\int_0^x dy\frac{1}{r^{(k_1k_2k_3)}+i\epsilon},\quad d^{(k_1k_2k_3)}=\frac{1}{2^4\pi^2}\int_0^1 dx\int_0^x dy\frac{(y-1)}{r^{(k_1k_2k_3)}+i\epsilon}.
\end{align}

\end{document}